\documentclass[
  aps,
  prl,
  reprint,
  longbibliography,
  superscriptaddress
]{revtex4-2}

\usepackage{amsmath,amssymb}
\usepackage{bm}
\usepackage{graphicx}
\usepackage{xcolor}
\usepackage{hyperref}
\usepackage{physics}
\usepackage{comment}
\usepackage[normalem]{ulem}
\usepackage{xurl}
\usepackage{orcidlink}
\usepackage{bbold}
\hypersetup{breaklinks=true}

\hypersetup{
  colorlinks=true,
  linkcolor=blue,
  citecolor=blue,
  urlcolor=blue,
  pdftitle={replica_transition_ising}
}

\definecolor{myteal}{HTML}{0F6E6C}

\newcommand{\ra}{\rightarrow}

\newcommand{\mc}[1]{\mathcal{#1}}

\newcommand{\unige}{
  Department of Theoretical Physics, University of Geneva,
  24 quai Ernest-Ansermet, 1211 Gen\`eve, Switzerland
}

\newcommand{\gqc}{
  Geneva Quantum Center, University of Geneva
}

\newcommand{\lptms}{
  Universit\'e Paris-Saclay, CNRS, LPTMS, 91405, Orsay, France
}
\usepackage{pdfpages}

\makeatletter
\AtBeginDocument{\let\LS@rot\@undefined}
\makeatother
\begin{document}

\title{Replica Transition at Measurement-Altered Ising Criticality}

\author{Kabir Khanna\:\orcidlink{0000-0001-9400-1597}}
\affiliation{\unige}
 \affiliation{\gqc}

\author{Sara Murciano\:\orcidlink{0000-0002-1638-5692}}
\affiliation{\lptms}

\author{Romain Vasseur\:\orcidlink{0000-0002-4636-4139}}
\affiliation{\unige}
\affiliation{\gqc}

\date{\today}

\begin{abstract}
Performing measurements on gapless quantum ground states can give rise to new universal phenomena, whose origins remain poorly understood even for the paradigmatic Ising critical point. Replica methods offer a natural framework to study such post-measurement ensembles, with the replica limit $Q\to 1$ describing physical Born-averaged observables. Here we show that, for the monitored critical Ising chain, the replicated post-measurement ensemble of states itself undergoes a phase transition precisely at the physically relevant replica limit $Q \ra 1$, preventing standard replica treatments. We use boundary conformal field theory (BCFT) methods to explain the underlying mechanism of this transition. Our methods yield a perturbative analytic calculation of the universal $\mc O(1)$ constant in the Shannon entropy of the measurement record, previously known only numerically. We also apply our tools to study  measurement-averaged spin correlators, providing a unified description of the universal properties of measurement outcomes and post-measurement states in the critical Ising chain.

\end{abstract}

\maketitle

\textit{Introduction --- }
Rapid progress in quantum 
simulators~\cite{RevModPhys.86.153,PRXQuantum.2.017003,DaleyQSimulations} has considerably broadened the paradigm of phases of matter. Measurements, which were once a passive means to probe a static many-body state, are now increasingly viewed as also a source of rich universal dynamics~\cite{PhysRevLett.128.060601, Ippoliti2022solvablemodelofdeep, choi_preparing_2023,PRXQuantum.4.010311, PRXQuantum.4.030322, PhysRevA.107.032215, Chan_2024, PhysRevX.14.041051, 10.21468/SciPostPhys.18.3.107, milekhin2024observable} and criticality~\cite{PhysRevB.98.205136, PhysRevB.100.134306, PhysRevX.9.031009, PhysRevX.10.041020, PhysRevLett.125.070606,PhysRevB.101.060301, PhysRevB.101.104301, PhysRevB.101.104302, hoke_measurement-induced_2023, koh_measurement-induced_2023, annurev:/content/journals/10.1146/annurev-conmatphys-031720-030658, Potter2022}, and as an efficient way to prepare quantum states~\cite{PhysRevLett.86.5188, PhysRevA.68.022312, briegel_measurement-based_2009, PhysRevLett.119.010504, PhysRevA.96.012302, PhysRevA.98.022332, PhysRevLett.122.090501,daniel2020computational, PhysRevX.14.021040, verresen2021efficiently, PRXQuantum.3.040337, iqbal_non-abelian_2024, foss2023experimental, PhysRevLett.131.200201, chen_nishimori_2025}. A key feature underlying such measurement-related phenomena is their intrinsic randomness. 
Measurement-driven phenomena, such as measurement-induced phase transitions~\cite{PhysRevB.98.205136,PhysRevX.9.031009}, are invisible to linear observables of the outcome-averaged density matrix and are instead revealed through quantities that retain information about the classical measurement record~\cite{PhysRevX.10.041020,PhysRevB.101.104301, PhysRevLett.129.200602, IppolitiKhemani2024}. 


Local measurements on gapless quantum matter can also radically reshape patterns of many-body entanglement~\cite{garratt2023measurements, PhysRevX.13.041042, sala24, liu24,PhysRevB.108.165120, PRXQuantum.4.030317, cheng2024universal, patil2024highlycomplexnovelcritical, Lin2023probingsign, khanna2025measurementinducedentanglementconformalfield, khanna2026universalstatisticsmeasurementinducedentanglement,eisler2026measurementinducedentanglementhamiltonian,rajabpour2026quantumsnapshotsrevealcompact, breach2026structuremeasurementinducedentanglementinfiniterandomness}.
Quantum critical ground states provide a controlled setting to explore universality from such measurement-driven phenomena~\cite{garratt2023measurements, PhysRevX.13.041042}. In such settings, one typically performs a ``weak" version of the more familiar projective measurement on the ground state while conditioning it on the measurement outcomes. Viewing such weak measurements as perturbations to the original state, one can use field theory and RG tools to analyze its fate~\cite{garratt2023measurements} in the same spirit as the Kane-Fisher treatment of impurities~\cite{PhysRevB.46.15233, PhysRevLett.68.1220}. In many cases, weak measurements, when relevant in the sense of renormalization group (RG), flow towards the \textit{projective limit}~\cite{kumar2026universalcrossoversweaklymonitoredquantum, Khanna:2026dcg}, where the measurement strength grows unbounded (with important exceptions discussed in Ref.~\cite{patil2024highlycomplexnovelcritical}). At the microscopic level, this fixed point may correspond to collapsing the measured degrees of freedom onto definite outcomes, leaving no remnants of the original critical state. While the resulting state may then be trivial at this fixed point, the \textit{approach} to the fixed point can still exhibit non-trivial universal measurement-altered correlations, and the ensemble \textit{at} the fixed point itself retains highly nontrivial universal information in the form of the Born distribution of the classical measurement record. In a recent work, we proposed a general framework for characterizing such projectively altered conformal critical points and their universal signatures~\cite{Khanna:2026dcg}. 
 \begin{figure}
    \centering
    \includegraphics[width=\linewidth]{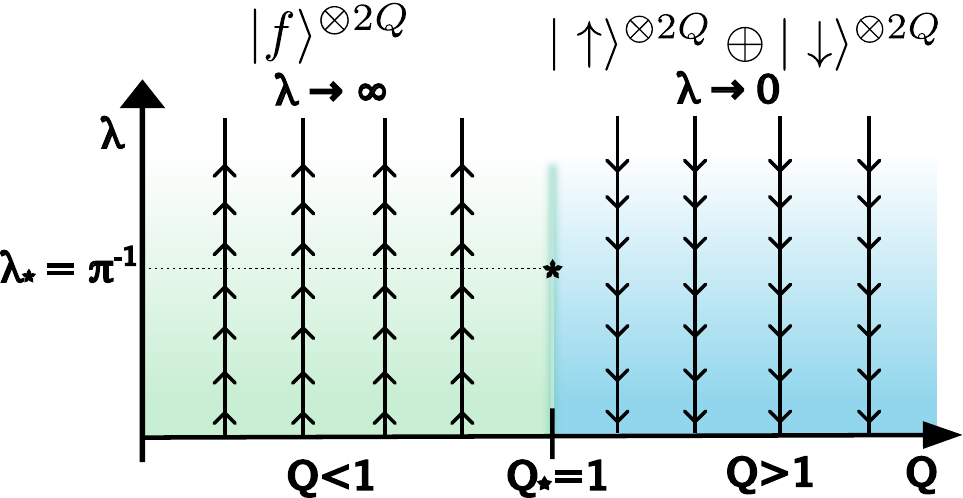}
    \caption{\textbf{Replica phase diagram.}
For $Q>1$, the BCC fugacity $\lambda$ is irrelevant and $\lambda\to0 $ in the IR, keeping the cat state $\ket{\uparrow}^{\otimes 2Q}\oplus \ket{\downarrow}^{\otimes 2Q}$ stable. Conversely, for $Q<1$, $\lambda$ becomes relevant and destabilizes the cat-state, driving the system toward the infrared fixed point $\lambda\to\infty$, characterized by the freely fluctuating boundary state $\ket{f}^{\otimes 2Q}$. At the singularity $Q_*=1$, the completeness condition $\sum_m \hat M_m^\dagger \hat M_m = \mathbb{I}$ fixes the coupling to $\lambda_*=\pi^{-1}$, corresponding to the no-defect point on the Oshikawa--Affleck defect line~\cite{OSHIKAWA1997533}.}
    \label{fig1maintext}
\end{figure} 

A simple field theory realization of this occurs when one weakly measures the order parameter in the ground-state of the Ising CFT. A straightforward microscopic realization corresponds to projectively measuring the lattice order parameter in the ground state of the critical transverse-field Ising chain in 1d, whose Born-distribution is simply given by the squared amplitudes of the critical ground-state wavefunction in the corresponding basis.
This distribution has been studied for longer than a decade through its Shannon and Rényi entropies~\cite{PhysRevB.80.184421,PhysRevLett.107.020402,PhysRevB.82.125455,PhysRevB.90.045424,PhysRevLett.111.017201,PhysRevLett.112.057203,PhysRevB.90.075132}. However, despite the Ising CFT being one of the best-understood critical theories, the universal information encoded in this distribution has remained a longstanding puzzle. In particular, for Rényi index $Q$, numerical works~\cite{PhysRevB.80.184421, PhysRevB.82.125455} find a phase transition detected by Rényi entropies as $Q$ is varied, with the transition occurring precisely \textit{ at} the Shannon $Q\ra 1$ point (see Fig.~\ref{fig1maintext}). From a measurement-induced criticality perspective, the location of the transition is striking since the Shannon $Q\ra 1$ point corresponds to averaging the entropy of outcomes by weighing them with their physical Born-probabilities. In the field theory too, the parameter $Q$ acts as the number of copies (``replicas") of the field, and the $Q \ra 1$ limit of the replica field theory precisely recovers the physical Born-ensemble. In other words, the Ising projective fixed point is exceptional since it sits \textit{at} the critical point of a \textit{replica transition}. As a result, universal quantities cannot be obtained by smoothly continuing calculations from either of the neighboring replica phases, presenting a major bottleneck since replica techniques are routinely used to extract observables in such measurement problems. Despite numerical works that established universal contributions to the Shannon entropy~\cite{PhysRevB.80.184421, PhysRevB.82.125455}, their analytical origin is still an open question. 
The seminal study of Cardy~\cite{Cardy1991} of the Ising model in a random boundary field provides an RG argument that explains a spontaneous ordering of the boundary depending on $Q$, and later mathematical works have supported this picture~\cite{DuminilCopin2023}.
However, explaining why this transition occurs and connecting it to a direct analytic characterization of the universal quantities associated with the critical Shannon point is still lacking.

In this Letter, we propose the underlying mechanism for this transition and give an analytic calculation of universal quantities associated with the critical Shannon point. We use tools from boundary conformal field theory (BCFT) and identify a replica `boundary-condition-changing' (BCC) operator that becomes marginal at the critical point, which then allows us to directly understand the critical point rather than performing an analytic continuation from either phase. As a result, we perturbatively evaluate the universal sub-leading constant in the Shannon entropy of the measurement record, previously only known numerically. 
We further apply our analysis to 
measurement averaged spin-correlations near the projective-fixed point, offering a holistic characterization of the universality of this phase transition.\\

\textit{Setup---} We consider the $c=1/2$ Ising conformal field theory (CFT) on a cylinder of circumference $L$ and height $\beta$, or equivalently, a massless Majorana field with left and right moving components $\psi_L$ and $\psi_R$, respectively, governed by the Euclidean action
\begin{equation}
\label{isingeuclideanaction}
    S_0 = \frac{1}{2\pi}\int d\tau dx \left[\psi_R(\partial_\tau - i\partial_x)\psi_R + \psi_L(\partial_\tau + i\partial_x)\psi_L \right], 
\end{equation}
with the corresponding partition function $Z_0=\tr e^{-\beta \hat H}=\int \mc D\psi_R\mc D\psi_L e^{-S_0[\psi_R,\psi_L]}$, where $\hat H$ is the Hamiltonian, $x\in[0,L)$ and $\tau\in[0,\beta]$ are spatial and imaginary-time coordinates, and $\beta$ is the inverse temperature. The above theory describes many physically relevant $\mathbb Z_2$ quantum critical points, such as in the transverse-field Ising chain~\cite{Sachdev_2011}, 1d topological $p$-wave superconductor (Kitaev chain)~\cite{ Kitaev_2001}, and Rydberg atom setups~\cite{Keesling2019}. For concreteness however, we will refer to the critical transverse-field Ising chain $\hat H = - \sum_i [\hat{\sigma}^z_i \hat{\sigma}^z_{i+1} +\hat{\sigma}^x_i] $ realization with the order parameter $\langle\hat \sigma^z\rangle $ setting our convention, with $\hat{\sigma}^z_i$ being the Pauli-Z operator. We fix the parity sector $P =\prod\hat{\sigma}^x_i =+1$, which corresponds to the boundary conditions $\psi_{L,R} (x+L)=-\psi_{L,R}(x)$ on the Majoranas. Our primary interest will be in the ground-state $\ket{\psi_0}$ of this model obtained by the imaginary time evolution $\ket{\psi_0}\bra{\psi_0}\propto\lim_{\beta\ra\infty}e^{-\beta\hat H}$. In the path integral representation, the ground state then lives on the $\tau = 0^{+,-}$ slices corresponding to the bra and the ket, respectively. 

We perform a weak measurement of the spin field $\sigma(x)$ on $\ket{\psi_0}$ with a measurement strength $\gamma \in[0,\infty)$, where the limit $\gamma\ra\infty$ corresponds to the projective fixed point. Formally, this measurement is implemented by the Kraus operator $\hat M_{m}\propto\exp\left[-\frac{\gamma}{4}\int dx( \sigma(x) - m(x))^2\right] $, where the proportionality constant is set by the completeness condition $\sum_m\hat M_m\hat M_m^{\dag} = \mathbb I$, and where $m(x)$ is to be thought of as the coarse-grained version of measurement outcomes that one obtains on the lattice. In the path-integral picture, this procedure introduces a perturbation along the $\tau=0$ time slice, which can be treated as a boundary upon folding. The output of this procedure is an ensemble of states $\ket{\psi_m} = \hat M_m \ket{\psi_0}/\sqrt{p_m}$ with Born probabilities $p_m \equiv Z_m/Z_0 =Z_0^{-1}\int \mc D\psi_R\mc D\psi_L e^{-S_0[\psi_R,\psi_L]}\hat M_m^2$. The central question of this work is to then characterize the universal features of this distribution in the low-energy theory. A way to do this is to introduce the replicated Born weight $p_m^Q$ and the associated replica partition function $Z(Q) = \sum_m p_m^Q$. The associated free-energies are of direct interest and are related to the $Q$-th Rényi Shannon entropy $R(Q)$ of the distribution via the relation $R(Q) = \frac{1}{1-Q}\log Z(Q)$, with $S = \lim_{Q \ra 1}R(Q) = -\sum_mp_m\log p_m$ giving the Shannon entropy of the measurement record. This entropy is known to scale as~\cite{PhysRevB.80.184421, patil2024highlycomplexnovelcritical, patil2025shannonentropymeasurementrecord}
\begin{equation}
\label{seffeqn}
S = -\sum_mp_m\log p_m = \alpha L - s_{\rm eff} +\dots,
\end{equation}
with $s_{\rm eff} = \log g_{\rm eff}$ a universal number (effective Affleck-Ludwig entropy~\cite{PhysRevLett.67.161}) that characterizes the effective boundary condition (or defect line) induced by the random measurement outcomes.

We will also use the same ensemble to study measurement-averaged moments of spin-correlations, 
\begin{equation}\label{eq:correlators}
    \mc C_k(x) = \sum_m p_m \left\langle\delta\sigma(0)\delta\sigma(x)\right\rangle_m^k,
\end{equation}
where $\delta \sigma = \sigma-\langle\sigma\rangle_m$, and where the expectation values are taken in the post-measurement state $\ket{\psi_m}$. \\

\textit{Replica Transition and the Born ensemble---} For positive integer $Q$, each factor of $Z_m$ in $Z_m^Q$ introduces a copy (replica) of the Ising CFT in the path integral, with the constraint that all of them share the same measurement record $m$ along $\tau=0$. In the Gaussian measurement protocol $\hat M_m$ we use in this work, one can proceed by integrating out the measurements, inducing an effective interaction between the $Q$ replicas described by the effective action $\tilde S_Q = \sum_{i=1}^Q S_{0,i} -\frac{\gamma}{Q}\int_{\tau=0} dx \sum_{b<c} \sigma_b(x)\sigma_c(x)$. Since the replica interaction $\sigma_a\sigma_b$ has scaling dimension $1/4$, a simple tree-level RG analysis suggests that the coupling grows under RG. The interaction favors alignment between replica spin fields along $\tau=0$, so one expects the boundary to flow to a conformal BC compatible with this constraint. In a single Ising BCFT, there exist \textit{three} conformal BCs, namely, the \textit{fixed} conformal (or Cardy) boundary conditions $\ket{\uparrow}$, $\ket{\downarrow}$, and the freely fluctuating $\mathbb Z_2$ symmetric \textit{free} BC $\ket{f}$. Accordingly, one expects all the replicas to flow to one of the two fixed BCs $\ket{\uparrow}$ or $\ket{\downarrow}$. While we find this to be the case for $Q>1$ (as we will see in the next section),
the stability of this order at the boundary depends on $Q$. To see this, we fold the replicas along $\tau=0$, obtaining $2Q$ copies of the Ising CFT on a half-plane $\tau<0$. Then, we note the existence of a BCC operator, labeled $\psi_{\uparrow\downarrow}$ (and its conjugate $\psi_{\downarrow\uparrow}$), at the fixed boundary, which when inserted at a point changes the boundary condition from $\ket{\uparrow}$ to $\ket{\downarrow}$ (see~\cite{supplement} for details). Since the interaction requires all replicas to agree, such a BCC must be inserted in the same way in all $2Q$ replicas. In the scaling limit, we interpret measurement outcome to flow to a piecewise fixed boundary conditions with BCC insertions at the domain wall positions. Summing over outcomes in the scaling limit then becomes a sum over the positions and number of these insertions, whose (normalized) partition function is
\begin{equation}
\label{replicabccexpansion}
    \frac{Z_Q(\lambda)}{Z_Q(0)} = \sum_{n=0}^{\infty}\lambda^{2n} \int_{x_1<\dots x_{2n}}\prod_{i=1}^{2n} dx_i \langle \psi_{\uparrow\downarrow}(x_1)\dots \psi_{\downarrow\uparrow}(x_{2n})\rangle^{2Q}_0, 
\end{equation}
where we introduce the BCC fugacity $\lambda$, and $\langle\cdots\rangle_0$ denotes a single-replica correlator with a fixed boundary condition. Periodic boundary conditions impose that the BCC insertions must come in pairs. Furthermore, the BCC $\psi_{\uparrow\downarrow}$ is a boundary \textit{Majorana} degree of freedom: it has a scaling dimension of $1/2$. Accordingly,  
the composite BCC placed at the boundary has dimension $(1/2)\times 2Q = Q$. This explains the underlying replica transition illustrated in Fig.~\ref{fig1maintext}: for $Q>1$, it is irrelevant, implying that the ordered boundary is stable, and in the IR $\lambda\to 0$. For $Q<1$, the number of BCCs proliferates, and they drive the system towards the freely fluctuating Ising BC $\ket{f}$ with $\lambda\to \infty$ in the infrared (IR). At $Q=1$ however, the BCC is \textit{exactly} marginal and the partition function can be solved exactly (see \cite{supplement}) since it is a quadratic in Majoranas at $Q=1$. The perturbation $\lambda$ at $Q=1$
then parametrizes a continuous line of conformal boundary conditions. To characterize this line, we can unfold the theory: the boundary perturbation then couples two Majorana chiral fields at $\tau=0$, and their matching condition can be expressed in terms of a scattering matrix that rotates the two components through the angle $\theta=\arctan(\pi\lambda)$. Bosonization identifies this family with the line of Dirichlet boundary conditions studied by Oshikawa and Affleck in~\cite{OSHIKAWA1997533}. More importantly, the physical Born ensemble \textit{selects} a particular point along this line, labeled $\lambda_*$. Indeed, the completeness relation of Kraus operators, $\sum_m \hat{M}_m^2=\mathbb{1}$, ensures that the Born-averaged (non-connected) post-measurement spin correlator retains the unmeasured decay $x^{-1/4}$, since it is linear in the density matrix and commutes with the measurement basis. Requiring the BCC description to reproduce this exponent fixes $\lambda_*=1/\pi$, or equivalently $\theta_*=\pi/4$, which corresponds precisely to the defect-free point of Oshikawa and Affleck (see~\cite{supplement}). Finally, as a result of this singularity at $Q=1$, the replica limit
must be taken {\em before} the thermodynamic limit in this problem by working directly in the $Q=1$ theory at finite-sizes, as both limits do not commute: the critical theory is described by $\lambda_*=1/\pi$ whereas the IR fixed points for $Q < 1$ and $Q>1$ correspond to $\lambda \to \infty$ and $\lambda \to 0$, respectively. 

Although our analysis focuses on the $\sigma^z$ basis here, our approach can be extended to the $\sigma^x$ basis by perturbing around the free, rather than fixed, Ising boundary condition. Indeed, the free boundary also supports a boundary magnetization operator of scaling dimension $1/2$, which can similarly drive a replica phase transition at $Q=1$, giving a dual phase diagram with a free phase for $Q>1$ and a cat-state phase for $Q<1$.

\textit{Shannon Entropy  ---} The replica transition discussed above manifests particularly clearly in the Rényi Shannon entropy of the measurement record. Numerical works~\cite{PhysRevB.80.184421, PhysRevB.82.125455} on the ground-state of the critical transverse field Ising chain find a universal sub-leading constant in the Rényi Shannon entropy $R(Q)$, which exhibits a sharp change across $Q=1$: $\log 2$ for $Q>1$ to $0$ for $Q<1$ with the $Q = 1$ value being reported as $0.2543925(5)$. These values have previously been associated with the Affleck-Ludwig (AL) boundary entropy~\cite{PhysRevLett.67.161} of the conformal boundary conditions in the underlying Ising BCFT. However, the precise manner in which this BCFT interpretation emerges remains unclear. This connection remains especially mysterious at $Q=1$, where the observed value does not coincide with the AL boundary entropy of any of the three standard Cardy boundary conditions. 

Concretely, for a BCFT on a cylinder of circumference $L$ and length $\beta$ with boundary conditions labeled $a$ and $b$, the boundary free-energy reads $F_{ab} = -\log Z_{ab} \underset{\beta/L\ra\infty}{\simeq} -\pi c\beta/(6L)-\log[\langle a|0\rangle\langle0|b\rangle]+o(1)$, where the sub-leading constants $s_{a,b} = \log g_{a,b}=\log\langle a,b|0\rangle$ are the universal AL boundary entropies~\cite{PhysRevLett.67.161}. 
The corresponding AL boundary entropies for the conformal Ising BCs are given as  $\log g_{\uparrow} = \log g_{\downarrow} = -\frac{1}{2}\log 2$ and $\log g_{f} = 0$~\cite{PhysRevLett.67.161}. The replica boundary state whose AL entropy we label $g_{Q}$ here manifests as the sub-leading constant in the Rényi Shannon entropy via the relation $R(Q)=(1-Q)^{-1}\log Z(Q) \underset{\beta/L\ra\infty}{\simeq}-c_QL/a + (1-Q)^{-1}\log g_{Q}$, where $c_Q$ is a non-universal constant. Taking the $Q \ra 1$ limit gives an ``effective" AL boundary entropy $s_{\mathrm{eff}} = \log g_{\mathrm{eff}}$~\cite{patil2024highlycomplexnovelcritical, patil2025shannonentropymeasurementrecord} as the sub-leading constant (see Eq.~\eqref{seffeqn}), which must coincide with the sub-leading constant of the Shannon entropy in Refs.~\cite{PhysRevB.80.184421, PhysRevB.82.125455}. 

The $Q\lessgtr 1$ values in Refs.~\cite{PhysRevB.80.184421, PhysRevB.82.125455} follow straightforwardly. Recall that since the composite BCC $\prod_{i=1}^{2Q}\psi^{(i)}_{\uparrow\downarrow}(x)$ (with $i$ indexing the replica number) is irrelevant along the boundary $\tau = 0$ for $Q>1$, the boundary must flow to a fixed boundary condition in all replicas. In particular, since the replica interaction does not pin the sign of the orientation, we have a replica superposition of the two fixed conditions on all replicas, namely $\ket{\Uparrow_Q}\oplus\ket{\Downarrow_Q} = \ket{\uparrow_1\dots\uparrow_Q}\oplus \ket{\downarrow_1\dots\downarrow_Q}$. The single-replica version of this is sometimes referred to as the ``cat" state BC in the literature and has been studied in other physical settings~\cite{PhysRevX.7.041048, PhysRevX.11.041059}. The evidence for this is also seen in the AL boundary entropy associated with this BC: $\log g_{Q>1} =\log\left[2.(g_{\uparrow})^{2Q}\right] = (1-Q)\log 2$, where the extra factor of $2$ is crucial to account for the additional degeneracy coming from the two possible orientations. This then returns the required sub-leading constant in the Shannon entropy, $(1-Q)^{-1}\log g_{Q>1}=\log 2$. The $Q<1$ follows similarly from the understanding that the composite BCC is relevant and proliferates to renormalize the boundary to the everywhere fluctuating $\mathbb Z_2$ symmetric BC on all replica, namely $\ket{f}^{\otimes 2Q}$, whose $\log g_{Q<1} = \log g_{f}^{2Q} = 0$, resulting in the required sub-leading constant for the $Q<1$ Rényi Shannon entropy. 

 \begin{figure}
    \centering
    \includegraphics[width=\linewidth]{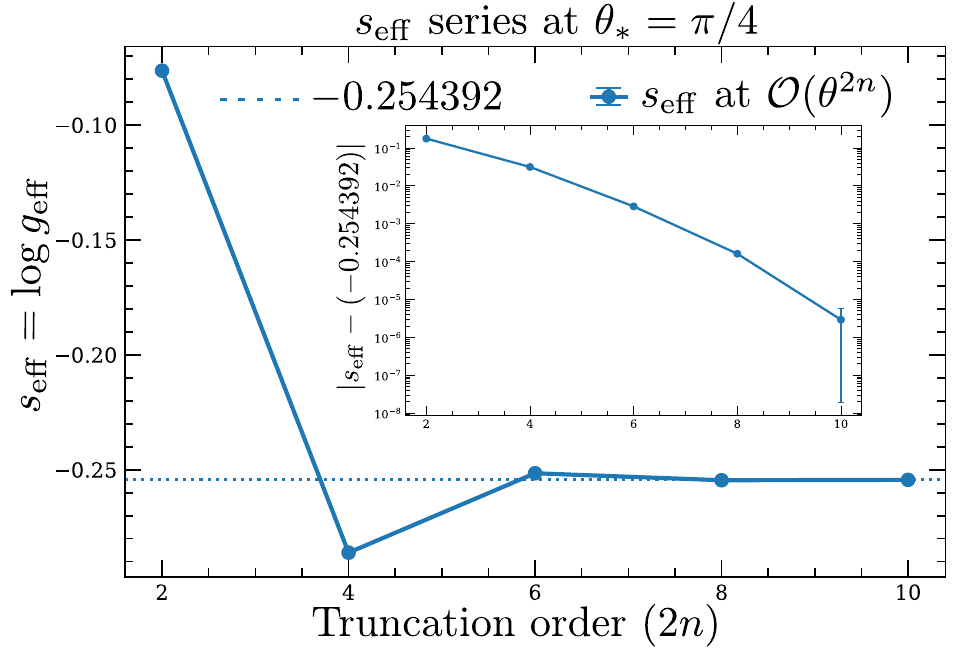}
    \caption{\textbf{Evaluation of $s_{\mathrm{eff}}$ in orders of $\theta^{2n}$}. Markers denote the partial sums of the series evaluated at $\theta_*=\pi/4$, truncated at order $\mathcal{O}(\theta^{2n})$. The horizontal dotted line indicates the reference value $s_{\mathrm{eff}}\simeq -0.254392$ reported in Ref.~\cite{PhysRevB.82.125455}. The inset shows the absolute deviation of the truncated series from this reference value on a logarithmic scale. The expansion coefficients are obtained by numerically evaluating the corresponding integrals.}
    \label{fig2maintext}
\end{figure} 

Clearly, a naive analytic continuation from $Q\lessgtr 1$ values fails at $Q=1$ due to the singularity at the transition. However, the BCC expansion at $Q=1$ of Eq.~\eqref{replicabccexpansion} can be used directly since it furnishes a finite-size description of the coarse-grained Born ensemble in the scaling limit, in orders of the number of BCC insertions. In particular, we can identify the universal content of $(2p_{\uparrow}^{Q})^{-1}\sum_m p_m^Q = Z_Q(\lambda_*)/Z_Q(0)$ in the scaling limit, where $p_{\uparrow}$ is the probability of the microscopic boundary condition $\ket{\uparrow\uparrow\dots\uparrow}$ (not to be confused with the conformal BC $\ket{\uparrow}$ which it ultimately renormalizes to in the IR). We emphasize that the above equality must be viewed as an equality between the \textit{universal} content of the two sides. The above follows from interpreting the insertion of BCCs as enumerating different measurement outcomes in the scaling limit. The factor of $2$ is due to the Ising $\mathbb{Z}_2$ symmetry relating $p_{\uparrow} = p_{\downarrow}$. Using this relation, we obtain~\cite{supplement}
\begin{equation}
    s_{\mathrm{eff}}=\log g_{\mathrm{eff}} = [\langle \log \mathrm{Det}\;G[X]\rangle_{\theta_* = \pi/4}]_{\mc O(1)}-\log 2, 
\end{equation}
where right hand side is evaluated at $\theta_* = \arctan\pi\lambda_* = \pi/4$, and where we must retain only the universal $\mc O(1)$ term. The $\log 2$ stems from the universal part of $\log p_{\uparrow}$ which is given as $2\log g_{\uparrow}=-\log 2$. The above expectation value is w.r.t the order-by-order BCC expansion Eq.~\eqref{replicabccexpansion} at the physical scattering point or Oshikawa-Affleck defect-free boundary condition $\theta_* =\arctan(\pi\lambda_*)= \pi/4$. At order $n$, denoting $X_n=[x_1,\dots,x_{2n}]$ such that $x_1<\dots<x_{2n}$ as a BCC configuration, $G[X_n]$ is the $2n\times 2n$ covariance matrix with entries corresponding to the finite-size BCC correlator $G_{ij} = \langle \psi_{\uparrow\downarrow}(x_i)\psi_{\downarrow\uparrow}(x_j)\rangle_0 =[(L/\pi)(\sin\pi (x_i-x_j)/L)]^{-1} $ with $G_{ii} =0$. The expectation value above is then understood as $\langle \log \mathrm{Det}\;G[X]\rangle =(Z_1(\lambda)/Z_1(0))^{-1}\sum_{n=1}^{\infty}\lambda^{2n}\int dX_n \;\mathrm{Det}G[X_n]\log \mathrm{Det}G[X_n]$. To obtain the required universal constant, one needs to evaluate the $\mc O(1)$ part of this series order by order in $\lambda^{2n}$. This is hard to do analytically for arbitrary order $n$. Nevertheless, we give expressions for this $\mc O(1)$ constant up to $\mc O(\lambda^4)$ analytically~\cite{supplement}. Moreover, we find that while the obtained series for the $\mc O(1)$ constant in $\lambda^2$ is poorly behaved, re-expressing the series in terms of the physical defect angle $\theta = \arctan(\pi\lambda)$ gives a series 
\begin{equation}
    \log g_{\mathrm{eff}}(\theta_*) + \log 2 = \theta_*^2 - \frac{\sqrt{3}}{\pi}\theta_*^4 + \mc O(\theta_*^6),
\end{equation}
which leads to $\log g_{\mathrm{eff}}$ to converge sufficiently close to the reported value in~\cite{PhysRevB.80.184421, PhysRevB.82.125455} already at $\mc O(\theta^6)$, where the series is evaluated at $\theta_* = \pi/4$. Higher-order terms require elaborate numerical evaluations with adaptive precision. We find that the series in $\theta$ converges rapidly, with coefficients decaying faster than factorially for the ones we can compute numerically (Fig.~\ref{fig2maintext}). Establishing the convergence of this series rigorously would be an interesting direction for future work.
Finally, we note that the universal constant $\log g_{\mathrm{eff}}$ in the dual ($\sigma^x$) basis differs from its $\sigma^z$ counterpart by $\log 2$, owing to the two-to-one mapping between spin configurations and domain walls under Kramers--Wannier duality~\cite{PhysRevB.82.125455}.

\textit{Measurement-Altered Correlations---} We now consider the Born-averaged correlators in Eq.~\eqref{eq:correlators}. We work directly with coarse-grained Born ensemble at $Q=1$, where the scaling dimension of the composite BCC $\psi_{\uparrow\downarrow}^{(1)}\psi^{(2)}_{\uparrow\downarrow}$ is 1. To evaluate such correlators, we first bring the spin fields $\sigma(0)$ and $\sigma(x)$ close to the uniform fixed boundary, retaining the leading operator contribution to the connected correlator. We then insert the BCCs that mimic the different measurement outcomes and sum over their configurations with the corresponding Born weights. The local boundary orientation also introduces a sign $s(x)=\pm 1$, since reversing the boundary orientation reverses the spin. The BCC insertions near $0$ and $x$ give the leading long-distance contribution proportional to $(s(0)s(x))x^{-2}$ (see~\cite{supplement}). As a consequence, the $k$-th moments contain $(s(0)s(x))^kx^{-2k}$. For even $k$, the local spin orientation is simply one, so it does not contribute. For odd $k$, the factor $s(0)s(x)$ remains and the Born average gives the unmeasured Ising decay $x^{-1/4}$. Putting everything together, we obtain
\begin{equation}
    \mc C_k(x)\sim\begin{cases}
        |x|^{-2k}\;\; \;\;\;\;\;\;k \in2\mathbb Z\\
        |x|^{-2k-1/4}\;\; k \in2\mathbb Z+1.\\
    \end{cases}
\end{equation}
Thus, the exponent grows linearly with $k$, with an additional $1/4$ for odd moments due to the critical correlations of the local boundary orientations. In particular, we find $C_1(x)\sim |x|^{-9/4}$ and $C_2(x)\sim |x|^{-4}$, consistent with the numerical results $C_1(x)\approx |x|^{-2.28}$ and $C_2(x)\approx |x|^{-4.02}$ found in~\cite{kumar2026universalcrossoversweaklymonitoredquantum}.

\textit{Summary---} In this Letter, we have established the physical mechanism explaining the replica transition at the Ising projective point. We identify its BCFT description and the BCC responsible for the transition, which is irrelevant for $Q>1$, relevant for $Q<1$, and exactly marginal at the physical Born average $Q=1$. This description yields a perturbative analytical calculation of the universal constant in the Shannon entropy, $s_{\rm eff}$, which was only known numerically before. We also apply our analysis to evaluate
the decay exponents of the even and odd moments of the connected spin correlators. Our results reproduce previous numerical estimates of the universal terms in the Shannon entropy of
critical Ising ground-state wavefunctions, providing a common explanation in terms of measurement-altered criticality. The expansion about the projective fixed point connects our framework to the replica-instanton approach developed for Luttinger liquids in Ref.~\cite{Khanna:2026dcg}. The Ising case, however, requires a distinct analysis because of the exactly marginal perturbation at $Q=1$. Beyond determining post-measurement averaged correlators, this analysis resolves longstanding questions about the universal contributions to the Shannon entropy.
Moreover, even though we mainly focus on Ising CFT, this approach only relies on the boundary description of the projective fixed point and the corresponding BCC operators. We therefore expect it applies to any CFTs whose measurements flow to such projective fixed points. For CFTs with richer operator content than Ising, identifying the correct BCC which determine the universal measurement-altered behavior is more challenging, but it can be done within the framework developed here. A broader open direction is to investigate measurement-altered criticality in higher spatial dimensions, where CFTs (and BCFTs) are less developed than in $1+1$ dimensions. In particular, it would be interesting to determine whether replica transitions occur at $Q=1$ in $2+1$ dimensions, and what universal behavior is captured by entropies and correlators. Finally, our approach could be the starting point also to evaluate the distribution of the measurement-induced entanglement and mutual information between spatially separated regions~\cite{khanna2025measurementinducedentanglementconformalfield,khanna2026universalstatisticsmeasurementinducedentanglement}.

 {\it Acknowledgments -- } This work was supported by the Swiss National Science Foundation (grant 10008234) and by the Foundation for the University of Geneva. 
 We thank Sam Garratt, Sarang Gopalakrishnan, Abhishek Kumar, Andreas Ludwig, Rushikesh Patil for insightful discussions and/or collaborations on related topics. The authors acknowledge the use of ChatGPT for help with analytically evaluating the $\mc O(1)$ contribution analytically at $\mc O(\theta^4)$ and polishing the manuscript. The numerical evaluation of the higher order terms in the $g_{\rm eff}$ series requires arbitrary precision numerics, and was performed using Claude Fable 5.1 and ChatGPT Sol 6.1. 
 

  {\it Note added -- } During the preparation of this manuscript, Ref.~\cite{rajabpour2026analyticalresultsshannonentropy} appeared on arXiv, reporting results for the Shannon entropy of the measurement record in the Ising chain in a basis dual to ours using a purely lattice approach. Our field-theoretic treatment elucidates the universal mechanism underlying this behavior and yields predictions for measurement-altered criticality beyond the Shannon entropy. The results on the $\log g_{\mathrm{eff}}$ expansion appear to agree.

\bibliographystyle{apsrev4-2}
\bibliography{apssamp}
\bigskip
\onecolumngrid
\newpage


\section*{Supplementary material}

# Supplemental Material

## Replica Transition at Measurement-Altered Ising Criticality

Kabir Khanna[1,2], Sara Murciano[3], and Romain Vasseur[1,2]
[1] *Department of Theoretical Physics, University of Geneva, 24 quai Ernest-Ansermet, 1211 Genève, Switzerland*
[2] *Geneva Quantum Center, University of Geneva*
[3] *Université Paris-Saclay, CNRS, LPTMS, 91405, Orsay, France*

### CONTENTS



## I. DETAILS OF THE BOUNDARY THEORY

In this section we present the details of the boundary theory in the IR upon weakly monitoring the order parameter ($\sigma$) field in the ground state $|\psi_0\rangle$ of the Ising model *at criticality*, described by the Euclidean action

$$S_0 = \frac{1}{2\pi}\int_0^\beta d\tau \int_0^L dx\,[\psi_R(\partial_\tau - iv\partial_x)\psi_R + \psi_L(\partial_\tau + iv\partial_x)\psi_L]\,, \tag{1}$$

where $v$ is the Fermi velocity that we set to unity by re-scaling coordinates, and $\psi_{R(L)}$ are the right (left) moving Majorana fermions. The partition function $Z_0 = \mathrm{tr}\, e^{-\beta\hat{H}} = \int \mathcal{D}\psi_R\mathcal{D}\psi_L e^{-S_0[\psi_R,\psi_L]}$ fixes the Hamiltonian with $\beta^{-1}$ setting the temperature. The ground state $|\psi_0\rangle$ is then obtained by taking the limit $\beta\to\infty$ in the above action: $|\psi_0\rangle\langle\psi_0| \propto \lim_{\beta\to\infty} e^{-\beta\hat{H}}$. The weak-monitoring of the $\sigma$ field on this ground state is performed via the Kraus operator

$$\hat{M}_m \propto \exp\left[-\frac{\gamma}{4}\int_{\tau=0} dx\,(m(x)-\sigma(x))^2\right], \tag{2}$$

where $\gamma$ is the measurement strength, with $\gamma\to\infty$ corresponding to the projective measurement limit, and $m$ is the coarse-grained measurement outcome. The proportionality constant is set by requiring the completeness condition $\sum_m \hat{M}_m\hat{M}_m^\dagger = 1$. The measurement operation on the ground state $|\psi_0\rangle$ results in an ensemble of states $|\psi_m\rangle$ indexed by their respective measurement outcomes, with corresponding Born probabilities $p_m$. These are related as

$$|\psi_m\rangle = \frac{\hat{M}_m|\psi_0\rangle}{\sqrt{p_m}}\ ,\ \ p_m = \langle\psi_0|\hat{M}_m^2|\psi_0\rangle. \tag{3}$$

In the Euclidean path integral representation, one can write

$$p_m = \lim_{\beta\to\infty}\frac{\mathrm{Tr}\left[e^{-\beta\hat{H}}\hat{M}_m^2\right]}{\mathrm{Tr}\left[e^{-\beta\hat{H}}\right]} = \frac{Z_m}{Z_0} = \frac{1}{Z_0}\int \mathcal{D}\psi_R\mathcal{D}\psi_L\ \hat{M}_m^2\exp\left[-S_0\right]. \tag{4}$$

Throughout this work, we will be interested in the Affleck-Ludwig Boundary entropy [1] (which is defined in Sec. II) and moments of spin-spin connected correlators, averaged over this ensemble of states. These observables are necessarily *non-linear* in the density matrix and require a replica trick to be used in order to calculate them. To see this, consider a simple non-linear observable $\langle\hat{O}\rangle^2$ averaged over measurement outcomes: $\overline{\hat{O}^2} = \sum_m p_m \langle\hat{O}\rangle_m^2$, where we take $\hat{O}$ to be some non-trivial operator in our theory. The replica trick can be then used as follows: we begin by introducing $Q$ copies of our system and write this average as

$$\overline{\hat{O}^2} = \sum_m \frac{\left[\langle\psi_0|\,\hat{M}_m\hat{O}\hat{M}_m\,|\psi_0\rangle\right]^2}{p_m} = \lim_{Q\to 1}\sum_m \left[\langle\psi_0|\,\hat{M}_m\hat{O}\hat{M}_m\,|\psi_0\rangle\right]^2\left[\langle\psi_0|\,\hat{M}_m^2\,|\psi_0\rangle\right]^{Q-2}. \tag{5}$$

The above is easily checked using Eq. (3). The idea is that two out of the $Q$ replicas of the system contain an insertion of the operator $\hat{O}$ while the rest $Q-2$ of them contain no insertion and contribute as the probability $p_m$. Therefore, in the replicated theory, one can re-cast each term in the above summation as an ordinary correlator that is easier to evaluate in principle. Finally, one must take the replica limit $Q\to 1$ which yields the desired result. In contrast, it is not too hard to repeat the exercise for linear observables and show that these are described entirely via the averaged density matrix $\bar{\rho} = \sum_m p_m\,|\psi_m\rangle\,\langle\psi_m|$. Viewing $\bar{\rho}$ as dephasing in the measurement basis, such observables in most cases erase any non-trivial effects of measurements.

To make progress with the replica trick, one typically sums over the coarse-grained measurement outcome field $m(x)$ to obtain an effective replica interaction. For simplicity, we demonstrate this with $Q$ replicas where we do not insert any observables to begin with. Consequently, we consider $Z(Q) = \sum_m p_m^Q$ and write this in the replicated field theory as

$$Z(Q) = \sum_m p_m^Q = \frac{1}{Z_0^Q}\int \mathcal{D}m(x)\int\prod_{i=1}^{Q}\mathcal{D}\psi_{R,i}\mathcal{D}\psi_{L,i}\;\hat{M}_{m,i}^2\exp[-S_{0,i}] \tag{6}$$

$$= \frac{1}{Z_0^Q}\int\prod_{i=1}^{Q}\mathcal{D}\psi_{R,i}\mathcal{D}\psi_{L,i}\exp\left[-\tilde{S}_Q\right], \tag{7}$$

where we have integrated over the Gaussian in the second equation, and where $\tilde{S}_Q$ is the effective replica action

$$\tilde{S}_Q = \sum_i S_{0,i} - \frac{\gamma}{Q}\int_{\tau=0} dx \sum_{b<c}\sigma_b(x)\sigma_c(x), \tag{8}$$

with $S_{0,i}$ being the $i$th copy of the free action (1), and $\sigma_i$ being the order parameter field in the $i$th replica. Under coarse-graining, the parameter $\gamma$ obeys the tree-level RG equation

$$\frac{d\gamma}{d\ell} = [d-2\Delta_\sigma]\gamma(\ell) = \frac{3}{4}\gamma(\ell) + \mathcal{O}(\gamma^2), \tag{9}$$

where $d=1$ the dimension in which the measurement perturbation acts, and $\Delta_\sigma = 1/8$ is the scaling dimension of the order parameter field $\sigma$. The factor of 2 enters due to the product of two fields in the interaction (8). Therefore, at least at the tree level, one expects the parameter $\gamma$ to grow under RG and the boundary ($\tau = 0^{+,-}$ corresponding to bra and ket respectively) to flow to the projective fixed point $\gamma\to\infty$. Formally speaking however, the replica dependent term at $\mathcal{O}(\gamma^2)$ can lead to an intermediate fixed point but the RG is not controlled (see Refs. [2, 3]). Furthermore, numerical evidence in Ref. [3] indicates that the RG indeed terminates at the projective fixed point with no intermediate fixed point in between.

One then naturally expects this boundary projective fixed point to be characterized by the Ising conformal boundary conditions [4]. Among the three Cardy boundary conditions in the Ising boundary conformal field theory (BCFT), namely free ($|f\rangle$), and two fixed ($|\uparrow\rangle$) and $|\downarrow\rangle$), the projective limit corresponds to the two fixed BCs. This is expected since we monitor the order parameter field and the replica interaction favors the alignment of spins, while not pinning the absolute sign/orientation. An infinitesimal boundary magnetic field can then select one of these sectors according to its sign, whereas in the absence of such a symmetry-breaking field, one must retain both the fixed BCs (we will elaborate more on this in Sec. II).

Numerical evidence suggests that the above tree-level analysis holds only for $Q>1$ and there occurs a boundary transition across $Q=1$ [5–8]. Instead, for $Q<1$, numerics in those works suggest that the boundary flows to $|f\rangle$ BC, while the physical $Q=1$ point remains elusive and does not seem to apriori correspond to any of the three Cardy BCs. In particular, the transition occurs precisely at the replica limit (Q=1), where Born-averaged observables

such as Eq. (5) are recovered, rendering the replica limit singular and necessitating a careful reconsideration of the standard replica approach. In the coming sections, we explain in detail the mechanism underlying this transition and show how the replica formalism can nevertheless be used to evaluate measurement-averaged observables and extract universal data analytically at the physical $Q = 1$ boundary theory.

### A. Perturbing about the Fixed Boundary Condition with BCCs

In order to explain the aforementioned transition, we will require the notion of boundary condition changing (BCC) operators in a BCFT [4]. A BCFT is characterized by a set of boundary operators and their respective scaling dimensions. Among these boundary operators, there exists a subset of BCC operators that implement a change between one conformally invariant BC to another when inserted at a point. The tower of scaling dimensions of these BCC operators are enumerated in boundary partition functions $Z_{ab}$ of the respective BCFT on a cylinder with boundaries $a$ and $b$ on either ends, where $a$ and $b$ are conformal BCs. See Ref. [9] for a good review of the basic ideas on the topic.

Here we consider the boundary partition function of the Ising BCFT with $a = \uparrow$ and $b = \downarrow$ on a strip of circumference $\beta$ and length $L$ [4],

$$Z_{\uparrow\downarrow}(q) = \chi_\epsilon(q) \underset{q\to 0}{=} q^{-1/48}\left[q^{1/2} + q^{3/2} + q^{5/2} + q^{7/2} + 2q^{9/2} + \dots\right], \tag{10}$$

where $q = e^{-\pi\beta/L}$, and where we have separated the $q^{-1/48}$ piece corresponding to the Casimir energy. $\chi_\epsilon(q)$ is the character corresponding to the $\epsilon$ (energy) primary in the Ising model. The remaining small-$q$ series then has a concrete interpretation: the powers of $q$ directly correspond to the BCC operator and its descendants, along with any degeneracies. In particular, the leading $q^{1/2}$ term is the most relevant BCC operator with a scaling dimension of $h = 1/2$ that changes $|\uparrow\rangle$ to $|\downarrow\rangle$ (and vice versa). We hereon label this BCC operator by the symbol $\psi_{\uparrow\downarrow}$ (and its conjugate partner $\psi_{\downarrow\uparrow}$).

We then imagine considering the $Q$ replicas of our theory, where the slices $\tau = 0^{+,-}$ can be taken to be in one of the fixed BC sectors, say $|\uparrow\rangle$. Since the upper half and lower half cylinders (corresponding to $\tau > 0$ and $\tau < 0$) are effectively disconnected, we can fold along $\tau = 0$ resulting in $2Q$ replicas of the theory on a semi-infinite cylinder $\tau < 0$. At the boundary of these replicas, we consider inserting the aforementioned $h = 1/2$ BCC, changing the BC from $|\uparrow\rangle \to |\downarrow\rangle$ at the inserted point. Such a BCC must be inserted simultaneously in all $2Q$ replicas at the corresponding point since the replica interaction (8) enforces all replicas to agree. Alternatively, the replica interaction required the replica fields $\sigma_a(x)$ to share the same fluctuating coarse-grained measurement outcome $m(x)$ before we integrated it out. Therefore, we can interpret the fluctuating coarse-grained measurement outcome $m(x)$ as configurations of this replicated (or composite) BCC operator $\Psi^{(Q)}_{\uparrow\downarrow}(x) = \prod_{i=1}^{2Q} \psi^{(i)}_{\uparrow\downarrow}(x)$. Since the scaling dimension of this operator is $(1/2) \times (2Q) = Q$, it is relevant for $Q < 1$, marginal for $Q = 1$, and irrelevant for $Q > 1$ at the boundary. Concretely, we introduce a parameter $\lambda$ which acts as the fugacity for these replicated BCC insertions, where $\lambda$ has the RG flow equation

$$\frac{d\lambda}{d\ell} = (1 - Q)\lambda, \tag{11}$$

at tree-level. This immediately gives us the phenomenology of the replica transition. Starting from a fixed BC, inserting the $h = 1/2$ BCC on the $\tau = 0$ slice is irrelevant for $Q > 1$, preserving the fixed boundary $|\uparrow\rangle$, while for $Q < 1$, the BCCs proliferate naturally resulting in the $\mathbb{Z}_2$ symmetric $|f\rangle$ BC. The $Q = 1$ case is marginal and needs to be treated separately. In fact, it is exactly marginal as we shall soon see. For a fixed $\lambda$, the resulting theory in terms of correlators of $\psi_{\uparrow\downarrow}$ and $\psi_{\downarrow\uparrow}$ is given as

$$\frac{Z_Q(\lambda)}{Z_Q(0)} = \sum_{n=0}^{\infty} \lambda^{2n} \int_{x_1 < \dots x_{2n}} \prod_{i=1}^{2n} dx_i \langle \psi_{\uparrow\downarrow}(x_1)\psi_{\downarrow\uparrow}(x_2)\dots\psi_{\downarrow\uparrow}(x_{2n})\rangle_0^{2Q}, \tag{12}$$

where $Z_Q(0)$ corresponds to the theory with no BCC insertions and a fixed boundary condition, and where we have an even number of BCCs inserted at $x_{i=1,\dots,2n}$ due to periodic boundary conditions. $\langle \dots \rangle_0$ is the expectation value in the theory without any BCC insertions and the fixed background. The above is analogous to the dual picture of boundary sine-Gordon in the free-boson BCFT where one expands about the IR Dirichlet fixed point [10]. In addition, the above expression is also consistent with the general picture proposed in Ref. [11] where one considers perturbing the boundary fixed point with BCCs.

Next, we can view this as a perturbation on the replicated action

$$S_Q(\lambda) = \sum_i S_{0,i}\Big|_{\text{fixed BC at } \tau=0} - \lambda \int_{\tau=0} dx\ \Psi^{(Q)}_{\uparrow\downarrow}(x), \tag{13}$$

with $\Psi^{(Q)}_{\uparrow\downarrow}(x) = \prod_{i=1}^{2Q} \psi^{(i)}_{\uparrow\downarrow}(x)$ being the composite BCC operator on the $2Q$ replicas after folding, and $S_0$ is the free action with a fixed BC imposed at $\tau = 0$. Furthermore, we can identify the individual $h = 1/2$ BCC with a chiral Majorana degrees of freedom restricted to the boundary. Indeed, an alternate way to interpret $\chi_\epsilon(q)$ (c.f. (10)) is that it enumerates the operators in the *chiral* Virasoro tower of the $\epsilon$ primary, of which the leading operator is realized as a chiral Majorana fermion that here acts as a BCC in the fixed BC background.

### B. Scattering Solution and Connection to the Orbifolded Boson

The above theory (13) is generically hard to solve for a given $Q$. However for $Q = 1$, the boundary perturbation is quadratic in the Majorana BCC and can be solved exactly as we shall detail in this section. The utility of this solution is two-fold. First, it will highlight that the parameter $\lambda$ is *exactly* marginal at $Q = 1$ with $\lambda$ by mapping it to the strength of the energy defect in the Ising Model as worked out by Oshikawa and Affleck [12]. Secondly, it will form a basis to extract the quantities that we wish to study in this work.

For $Q = 1$, we have the action

$$S_1(\lambda) = \frac{1}{2\pi} \sum_{i=1,2} \int_0^\infty d\tau \int_0^L dx\ [\psi_{R,i}(\partial_\tau - i\partial_x)\psi_{R,i} + \psi_{L,i}(\partial_\tau + i\partial_x)\psi_{L,i}] - \lambda \int_{\tau=0} dx \psi^{(1)}_{\uparrow\downarrow}(x)\psi^{(2)}_{\uparrow\downarrow}(x), \tag{14}$$

At $\lambda = 0$, the fixed boundary condition (both $|\uparrow\rangle$ and $|\downarrow\rangle$) in terms of Majoranas were considered in Ref. [13] where the authors considered the Ising model on the semi-infinite half plane. A simple conformal map to the cylinder needs to be done in order to get the corresponding boundary condition on a cylinder (strip), which can be worked out to be [14]

$$\psi_{R,i}(x, 0) = \psi_{L,i}(x, 0). \tag{15}$$

The above boundary condition at $\lambda = 0$, along with the fact that $\psi_R$ and $\psi_L$ are functions of $x + i\tau$ and $x - i\tau$ only, implies we can "unfold" the two semi-infinite half cylinders onto two infinite cylinders by defining a single chiral field $\chi_i$ on the entire cylinder $\tau \in (-\infty, \infty)$. Concretely, we can define

$$\chi_i(x, \tau) = \begin{cases} \psi_{R,i}(x, \tau) & \tau > 0 \\ \psi_{L,i}(x, -\tau) & \tau < 0. \end{cases} \tag{16}$$

It is then natural to identify $\sqrt{i}\chi_i(x, 0) \simeq \psi^{(i)}_{\uparrow\downarrow}(x, 0)$ up to normalization and the BCC label, where we define $\chi_i(x, 0) = 2^{-1}(\chi_i(0^-) + \chi_i(0^+))$. The interacting action then becomes

$$S_1(\lambda) = \frac{1}{2\pi} \sum_{i=1,2} \int_{-\infty}^\infty d\tau \int_0^L dx\ [\chi_i(\partial_\tau - i\partial_x)\chi_i] - i\lambda \int_{\tau=0} dx \chi_1(x)\chi_2(x), \tag{17}$$

where the $\sqrt{i}$ factor was included in order to get a Hermitian interaction. The above is a fairly standard free-fermion system with a quadratic impurity along the $\tau = 0$ boundary (see Ref. [14] for a realization of this in a different context). The consequence of this impurity is to simply produce a phase shift across $\tau = 0$. Concretely, if we define a two-component chiral fermion $\boldsymbol{\chi} = (\chi_1\ \chi_2)^T$, the solution to the above is given in terms of the scattering matrix

$$\boldsymbol{\chi}(0^+) = \mathcal{R}(\lambda)\boldsymbol{\chi}(0^-), \quad \mathcal{R} = \begin{bmatrix} \cos 2\theta & \sin 2\theta \\ -\sin 2\theta & \cos 2\theta \end{bmatrix}, \tag{18}$$

where $\theta = \arctan(\pi\lambda)$. Thus, $\lambda$ is indeed independent of any scales in the problem and is exactly marginal, parametrizing a continuous line of conformal boundary conditions. Concretely, it coincides with the Dirichlet line of Oshikawa and Affleck induced by inserting an energy defect at a single bond. To see this, we bosonize the two species of Majoranas, namely, $(\psi_{R,1}, \psi_{L,1})$ and $(\psi_{R,2}, \psi_{L,2})$ by defining the chiral complex fermions $\Psi_L = (\psi_{L,1} + i\psi_{L,2})/\sqrt{2}$ and $\Psi_R = (\psi_{R,1} - i\psi_{R,2})/\sqrt{2}$. The opposite sign in the definitions between $\Psi_L$ and $\Psi_R$ will be clear in a moment.

With this definition, we can re-write the above scattering relation in terms of the chiral fermions as $\Psi_R = e^{2i\theta}\Psi_L^\dagger$. Bosonizing the chiral fermions using $\Psi_R \sim e^{i\phi_R}$ and $e^{i\phi_L} \sim \Psi_L$, the scattering relation results in $\varphi = (\phi_R+\phi_L)/2 = \theta$, where $\varphi$ is the bosonic field in the conventions of Oshikawa and Affleck [12]. Therefore, the scattering angle $\theta$ directly corresponds to the Dirichlet line of Oshikawa and Affleck where one pins the bosonic field $\varphi$ along the defect $\tau = 0$. One could have instead used the defining chiral fermions to be $\Psi'_R = \Psi_R^\dagger$ and $\Psi'_L = \Psi_L$. It is easy to check that this would then result in the Neumann line for the dual bosonic field in the conventions of Oshikawa and Affleck. However, as per their original bosonic field conventions, we are along the Dirichlet line. This identification can be verified by matching scaling dimensions in the fermionic and bosonic descriptions. A particularly simple check is provided by the $\lambda = 0$ limit: the fixed BCs $|\uparrow\uparrow\rangle$ and $|\downarrow\downarrow\rangle$ correspond to the Dirichlet value $\varphi_0 = 0$ in their conventions, namely, $|\uparrow\uparrow\rangle \equiv D(0^+)$ and $|\downarrow\downarrow\rangle \equiv |D(0)^-\rangle$. Varying $\lambda$ should then move the boundary condition continuously away from this point.

### C. Fixing Marginal Perturbation Strength

The marginal parameter $\lambda$ in our problem is not arbitrary, but it is constrained by the completeness relation $\sum_m M_m^\dagger M_m = \mathbb{I}$. Since the spin operator $\sigma_i^z$ commutes with the measurement operator, its Born-averaged two-point correlator must decay in the same manner as in the unmeasured ground state, that is,

$$\sum_m p_m \langle \sigma^z(i)\sigma^z(j)\rangle_m = \langle \sigma(i)\sigma(j)\rangle \sim |i-j|^{-1/4}, \tag{19}$$

where $\sigma^z \simeq a^{1/8}\sigma$ is the lattice Pauli-$Z$ operator and $a$ is the UV cutoff. We can then use the above constraint to fix $\lambda$ to a specific value along the Oshikawa-Affleck defect line.

For this particular $\lambda = \lambda_*$, we propose the identification

$$\left.\frac{Z_Q(\lambda)}{Z_Q(0)}\right|_{\lambda=\lambda_*} = \frac{\sum_m p_m^Q}{2p_\uparrow^Q}, \tag{20}$$

where $p_{\uparrow(\downarrow)}$ is the probability of obtaining the microscopic $|\uparrow\uparrow\ldots\uparrow\rangle$ $(|\downarrow\downarrow\ldots\downarrow\rangle)$ boundary condition, which in the scaling limit corresponds to the two fixed BCs $|\uparrow\rangle$ $(|\downarrow\rangle)$ given by the $\lambda = 0$ theory. The factor of two follows from the $\mathbb{Z}_2$ symmetry, which ensures $p_\uparrow = p_\downarrow$. The identification Eq. (20) is motivated by interpreting the set of measurement outcomes to flow to the set of piecewise fixed boundary conditions with the changes being represented with BCC insertions. The BCC expansion in Eq. (12) therefore provides a continuum description of the measurement ensemble, organized as per the number of boundary-condition changes. We stress that the above equality should be read as an equality between the universal content of the two sides, rather than an exact equality at the microscopic level. Consequently, the universal content of the $Q = 1$ measurement ensemble is captured by the BCC expansion evaluated at $\lambda = \lambda_*$. In this field theoretic description, the constraint in Eq. (19) can be expressed as

$$\frac{Z_1(0)}{Z_1(\lambda)}\sum_{n=0}^{\infty}\lambda^{2n}\int_{x_1<\cdots<x_{2n}}\prod_{i=1}^{2n}dx_i\ \langle \Psi^{(1)}_{\uparrow\downarrow}(x_1)\Psi^{(1)}_{\downarrow\uparrow}(x_2)\ldots\Psi^{(1)}_{\downarrow\uparrow}(x_{2n})\sigma(0)\sigma(x)\rangle_0 \sim x^{-1/4}, \tag{21}$$

where $\Psi^{(1)}_{\uparrow\downarrow}(x) = \psi^{(1)}_{\uparrow\downarrow}(x)\psi^{(2)}_{\uparrow\downarrow}(x) \simeq i\chi_1(x)\chi_2(x)$ is the composite BCC inserted at the boundary of the $Q = 1$ theory along the bra and ket. The left-hand-side of the above equation can then be easily evaluated since the BCCs are Majoranas with their correlators obey Wick's theorem at the boundary. We begin by first considering $Z_1(\lambda)/Z_1(0)$ without any insertions of the spin-field. While this theory was already solved in the previous section, it is instructive to solve it again directly at the level of partition functions. Using Wick's theorem, we can write

$$\frac{Z_1(\lambda)}{Z_1(0)} = \sum_{n=0}^{\infty}\frac{\lambda^{2n}}{(2n)!}\int\prod_{i=1}^{2n}dx_i\ \mathrm{Det}\ G(x_1,\ldots,x_{2n}), \tag{22}$$

where $\mathrm{Det}G(x_1,\ldots,x_{2n})$ is the Majorana propagator matrix determinant with $G_{ij} = \langle\chi(x_i)\chi(x_j)\rangle_0 = d_c(x_i-x_j)^{-1}$ and $G_{ii} = 0$, where $d_c(x) = (L/\pi)\sin(\pi x/L)$ is the chord-length on a system of size $L$. The above follows from two facts: first, the standard Wick result $\langle\prod_{i=1}^{2n}\chi_1(x_i)\rangle = \mathrm{Pf}(G)$, and second, $\mathrm{Pf}^2[...] = \mathrm{Det}[...]$, since we have two sets of Majoranas at the boundary. Eq. (22) is precisely the expansion of a Fredholm determinant, leading to $Z_1(\lambda) = Z_1(0)\mathrm{Det}[1+\lambda\hat{G}]$, where we use $\hat{G}$ to denote the infinite-dimensional continuum version of $G(x,y)$,

such that $\langle x|\hat{G}|y\rangle = G(x,y)$. Furthermore, due to translational invariance, $\hat{G}$ is diagonal in momentum space with $G(k) = -i\pi\,\mathrm{sgn}(k)$, giving

$$\log\frac{Z_1(\lambda)}{Z_1(0)} = \int_{-\Lambda}^{\Lambda}\frac{L}{2\pi}dk\,\log[1+i\pi\lambda\,\mathrm{sgn}(k)], \tag{23}$$

where we have used the identity $\log[\mathrm{Det}[\dots]] = \mathrm{Tr}\log[\dots]$, and then converted the sum $\sum_k \to (2\pi/L)^{-1}\int dk$. Simplifying the above integral gives $Z_1(\lambda) = Z_1(0)\left[1+\lambda^2\pi^2\right]^{L\Lambda/(2\pi)} = Z_1(0)[1+\lambda^2\pi^2]^{L/(2a)}$, where $\Lambda = \pi/a$ is the momentum UV cutoff. Furthermore, it is useful to think of each term in Eq. (22) as defining a partition function of $2n$ BCC insertions where we sum over all possible configurations $X_n = (x_1,\dots,x_{2n})$. Alternatively, these are the coarse-grained Born-weights for the coarse grained measurement outcomes indexed by the set $\mathcal{X} = \{X_n|\ n\in\mathbb{Z}\}$. Accordingly, for an observable $O[X_n]$, we can define

$$\langle O\rangle_\lambda = \left(\frac{Z_1(\lambda)}{Z_1(0)}\right)^{-1}\left[\sum_{n=0}^{\infty}\frac{\lambda^{2n}}{(2n)!}\int\prod_{i=1}^{2n}dx_i\ \mathrm{Det}G[X_n]O[X_n]\right] \tag{24}$$

Following this, the spin-spin correlations can be evaluated as follows. The spin operator satisfies $\sigma|_{\tau=0,\lambda=0} = \pm 1$ (up to normalization) depending on the orientation of the fixed boundary. Consequently, for a given BCC configuration $X_n$ the product $\sigma(0)\sigma(x)$ then measures the parity of the number of BCCs within $(0,x)$. A convenient way to deal with (21) is to then introduce a function

$$\zeta_{x,y}(z) = \begin{cases}-1 & z\in[x,y]\\ +1 & z\notin[x,y]\end{cases}, \tag{25}$$

which records the parity of the number of BCCs between $(x,y)$ and write,

$$\langle\sigma(x)\sigma(0)\rangle_\lambda = \frac{Z_1(0)}{Z_1(\lambda)}\sum_{n=0}^{\infty}\lambda^{2n}\int_{x_1<\dots<x_{2n}}\prod_{i=1}^{2n}dx_i\ \mathrm{Det}G[X_n]\left(\prod_{i=1}^{2n}\zeta_{0,x}(x_i)\right) = \frac{\mathrm{Det}[1+\lambda\zeta_{0,x}\hat{G}]}{\mathrm{Det}[1+\lambda\hat{\mathrm{G}}]}, \tag{26}$$

where $\zeta_{0,x}$ is thought of as a diagonal $L\times L$ matrix with entries being $-1$ between the sites $(0,x)$ and $+1$ otherwise. Next, we can define the projector $P_x$ onto the region $(0,x)$ as being $P_x = (\mathbb{I}-\zeta_{0,x})/2$. This then gives

$$\frac{\mathrm{Det}_L[1+\lambda\zeta_{0,x}\hat{G}]}{\mathrm{Det}_L[1+\lambda\hat{G}]} = \mathrm{Det}_L[1-2\lambda\hat{G}(1+\lambda\hat{G})^{-1}P_x] = \mathrm{Det}_x\left[1-2\lambda\hat{G}(1+\lambda\hat{G})^{-1}\right], \tag{27}$$

where we have used $\mathrm{Det}[1+UV] = \mathrm{Det}[1+VU]$ and the notation $\mathrm{Det}_x[\dots]$ is understood as taking the determinant only within the $(0,x)$ block of the matrix. Within the $(0,x)$ block, the matrix $1-2\lambda\hat{G}(1+\lambda\hat{G})^{-1}$ is translationally invariant (Toeplitz) and can be simplified as

$$\mathrm{Det}_x\left[1-2\lambda\hat{G}(1+\lambda\hat{G})^{-1}\right] = \mathrm{Det}_{(0,x)}\left[\int_{-\Lambda}^{\Lambda}\frac{dk}{2\pi}e^{ik(x_1-x_2)}\frac{1+i\pi\lambda\,\mathrm{sgn}(k)}{1-i\pi\lambda\,\mathrm{sgn}(k)}\right], \tag{28}$$

where $x_1,x_2\in(0,x)$. It is instructive to express $(1+i\pi\lambda\,\mathrm{sgn}(k))(1-i\pi\lambda\,\mathrm{sgn}\,k)^{-1} = e^{2i\theta\,\mathrm{sgn}(k)}$, where we recover the phase $\theta = \arctan(\pi\lambda)$ we obtained in the previous section. This form helps us apply the Fisher Hartwig conjecture [15, 16]. It (loosely) states that for a Toeplitz matrix $T_\ell$ of size $\ell\times\ell$ with Fourier coefficients $c(k)\equiv c(e^{i\varphi}) = b(e^{i\varphi})\prod_{r=1}^{R}t_{\beta_r}(\varphi-\varphi_r)u_{\alpha_r}(\varphi-\varphi)$, with $t_\beta(\varphi) = \exp(i\beta(\varphi-\pi))$, $0<\varphi<2\pi, u_\alpha(\varphi) = (2-2\cos\varphi)^\alpha$, $\mathrm{Re}[\alpha] > -1/2$, and where $b(e^{i\varphi})$ is a sufficiently smooth function. $R$ is the number of discontinuities and divergences in $t_\beta(\varphi), u_\alpha(\varphi)$, the asymptotic behavior of the determinant is give by $\mathrm{Det}[T_\ell] \underset{\ell\to\infty}{\sim} G[b]^\ell \ell^{\sum_r \alpha_r^2-\beta_r^2}E$, where $E$ is a constant pre-factor independent of $\ell$. In our case, we only have a discontinuous piece (with $R=1$) with $\beta = (2/\pi)\arctan(\pi\lambda)$, leading to the scaling

$$\langle\sigma(0)\sigma(x)\rangle \sim \mathrm{Det}_x[1-2\lambda\hat{G}(1+\lambda\hat{G})^{-1}] \underset{x\to\infty}{\sim} x^{-2\Delta_\sigma(\lambda)}, \tag{29}$$

with $2\Delta_\sigma(\lambda) = \left[\frac{2}{\pi}\arctan(\pi\lambda)\right]^2$. In order to have $\langle\sigma(0)\sigma(x)\rangle\sim x^{-1/4}$, we therefore have the $Q=1$ marginal fugacity to be fixed to the value

$$\lambda_* = \pi^{-1}. \tag{30}$$

In the bosonic language, this corresponds to the no-defect Dirichlet boundary condition $\theta_* = \varphi_0 = \pi/4$ [12]. This is to be expected since we impose the $Q=1$ Born ensemble to return the scaling of the "clean" Ising CFT without any defect. This therefore also provides a non-trivial consistency check on the connection with the Dirichlet line of Oshikawa and Affleck.

## II. EFFECTIVE AFFLECK-LUDWIG BOUNDARY ENTROPY CALCULATION

In this section we will detail the calculation of the "effective" Affleck-Ludwig (AL) boundary entropy, or alternatively, the $\mathcal{O}(1)$ constant in the Shannon entropy of the measurement record in the critical Ising chain. We begin with a brief introduction.

Consider a 1+1D CFT with central charge $c$ on a strip of circumference $L$ and length $\beta$, with boundaries having conformal boundary conditions labeled as $a$ and $b$ respectively. We can quantize this system in the "closed channel", that is, by considering Euclidean time propagation along $\beta$. The resulting partition function in the limit $\beta/L \to \infty$ has the expansion [1]

$$F_{ab} = -\log Z_{ab} \underset{\beta/L\to\infty}{\simeq} \frac{-\pi c\beta}{6L} - \log\langle a|0\rangle - \log\langle 0|b\rangle + o(1), \tag{31}$$

where $s_{a,b} = \log g_{a,b} = \log\langle a,b|0\rangle$ are the universal Affleck-Ludwig boundary entropies associated with the two conformal boundaries $a$ and $b$ respectively, with $|0\rangle$ being the vaccum state and $|a,b\rangle$ being Cardy states [4] associated with the boundaries $a$ and $b$.

We can then connect the above formalism to our measurement problem. The measurement record $m$ defines a probability ensemble $\{m, p_m\}$ with $\sum_m p_m = 1$. We can then define the $Q$ replica partition function

$$Z(Q) = \sum_m p_m^Q, \tag{32}$$

where $Q$ can be thought of as inverse temperature by writing $p_m = e^{-E_m}$ with $E_m = -\log p_m$ being energies associated to the outcome $m$. The above definition directly connects to the Rényi Shannon-entropy of the measurement record,

$$R(Q) = \frac{1}{1-Q}\log\sum_m p_m^Q. \tag{33}$$

The Shannon entropy $S$ of the measurement record is obtained from the $Q \to 1$ limit as

$$S \equiv \lim_{Q\to 1} R(Q) = -\partial_Q \log Z(Q)|_{Q=1} = -\sum_m p_m \log p_m. \tag{34}$$

The field theory construction introduced in Sec. I allows us to cast the above partition function (32) into a boundary partition function: since measurements act on the $\tau = 0$ slice on all $Q$ replicas, folding along it turns $\tau = 0$ into a boundary with measurement outcome $m$ on all the (now $2Q$) replicas determining the boundary condition. Summing over measurement outcomes (as in Eq. (7)) then leads to an effective $2Q$ replica partition function $Z(Q)$ with a boundary along $\tau = 0$. The boundary after summing over outcomes $m$ can then be characterized by a $Q$-dependent Affleck-Ludwig $g$ value, denoted by $g_Q$. Concretely, since we are interested in the ground state limit $\beta/L \to \infty$ and scaling limit $L/a \to \infty$ with $a \ll L \ll \beta$, we can write $\log Z(Q) = -c_Q.(L/a) + \log g_Q + o(1)$ in the same way as Eq. (31), where note that the leading extensive term here is due to the boundary free-energy (unlike in Eq. (31) where the leading term is due to the bulk free-energy). Therefore, in this limit,

$$R(Q) = \frac{c_Q}{Q-1}\frac{L}{a} + \frac{\log g_Q}{1-Q} + o(1), \tag{35}$$

where the required sub-leading constant in the Shannon entropy is identified with $(1-Q)^{-1}\log g_Q$. In the $Q = 1$ case, completeness $\sum_m p_m = 1$ implies $\log g_1 = 0$ and so the physically interesting quantity is instead supplied by the sub-leading universal constant in the Shannon entropy, which can be expressed by taking the $Q \to 1$ limit in Eq. (35), giving

$$S = \alpha L - s_{\text{eff}} + o(1), \tag{36}$$

where $s_{\text{eff}} = \log g_{\text{eff}} = \partial_Q \log g_Q|_{Q=1}$ is referred to as the "effective" Affleck-Ludwig boundary entropy [2, 17].

In the 1d quantum Ising chain, the transition in the replica number $Q$ can be characterized by the Affleck-Ludwig boundary entropy. In particular, Refs. [5, 6] find the sub-leading $O(1)$ constant in the Shannon entropy to be:

$$\frac{\log g_Q}{1-Q} = \begin{cases} \log 2 & Q > 1 \\ 0.2543925(5) & Q \to 1 \\ 0 & Q < 1 \end{cases}, \tag{37}$$

where for $Q \to 1$, the above (is negative of) the effective Affleck Ludwig boundary entropy, that is, $-s_{\text{eff}}$ (c.f. (36)). Using the understanding of the boundary transition developed in Sec. I, we explain these values. In particular, we give an analytic derivation of the effective Affleck-Ludwig boundary entropy $s_{\text{eff}}$, which was previously only known numerically.

Starting with $Q > 1$, the composite BCC $\Psi^{(Q)}_{\uparrow\downarrow}$ is irrelevant and the boundary is in the ordered phase. Since the replica interaction does not favor a specific orientation, the boundary condition is formally a direct sum of both possible orders over all replicas, namely, $|\Uparrow_Q\rangle \oplus |\Downarrow_Q\rangle = |\uparrow_1\uparrow_2\uparrow \dots \uparrow_{2Q}\rangle \oplus |\downarrow_1\downarrow_2 \dots \downarrow_{2Q}\rangle$. Using $g_{\text{fixed}} = g_\uparrow = g_\downarrow = 1/\sqrt{2}$ [1], we have

$$g_Q = 2.(g_{\text{fixed}})^{2Q} = 2^{1-Q}, \tag{38}$$

which results in $(1-Q)^{-1} \log g_Q = \log 2$ for $Q > 1$. Notice that the factor of 2 due to the degeneracy between $|\Uparrow_Q\rangle$ and $|\Downarrow_Q\rangle$ is essential. For $Q < 1$, the BCCs $\Psi^{(Q)}_{\uparrow\downarrow}$ proliferate, causing the boundary to fluctuate everywhere, resulting in the free boundary condition factorized over all replicas, namely $|f\rangle^{\otimes 2Q}$. Since $g_Q = g_f^{2Q} = 1$ [1], it results in $(1-Q)^{-1} \log g_Q = 0$ identically for all $Q < 1$, as required.

The non-trivial $Q = 1$ case can be dealt with by using the identification Eq. (20), which leads to

$$\lim_{Q\to 1} \partial_Q \log \frac{Z_Q(\lambda)}{Z_Q(0)} = \sum_m p_m \log p_m - \log p_\uparrow, \tag{39}$$

The above equation poses a subtlety. For fixed $Q \neq 1$, the IR drives the BCC fugacity $\lambda$ to different fixed points on the two sides of $Q = 1$: $\lambda \to 0$ for $Q > 1$ and $\lambda \to \infty$ for $Q < 1$, whereas at $Q = 1$ the perturbation is exactly marginal and remains at a finite value $\lambda = \lambda_*$. Consequently, one cannot first replace the replicated boundary theory by its $Q \neq 1$ IR fixed point and then analytically continue to $Q = 1$. Indeed, it would be incorrect to analytically continue the result $g_{Q>1} = 2^{1-Q}$ to $Q = 1$ in order to obtain $s_{\text{eff}}$. However, one can use the BCC expansion Eq. (12) since while it is a continuum scaling description, it may be evaluated at large but finite $L/a$, and one may then organize the series in decreasing powers of $L/a$. Concretely, we may therefore first differentiate Eq. (12) w.r.t $Q$ at $Q = 1$, and only then subsequently extract the universal $\mathcal{O}(1)$ contribution in the limit $L/a \to \infty$, order by order in $\lambda^2$. This results in

$$\langle \log \text{Det}G[X]\rangle_{\lambda=\lambda_*} + [\partial_Q \log \lambda(Q, L)]\Big|_{Q=1} \langle N[X_n]\rangle_{\lambda=\lambda_*} = \sum_m p_m \log p_m - \log p_\uparrow, \tag{40}$$

where we have made explicit the $Q$ and $L$ dependence of the coupling constant $\lambda$. $\langle N[X_n]\rangle_\lambda$ is the average number of BCC insertions with $N[X_n] = 2n$ for each configuration $X_n$ in the expansion Eq. (24). Using $Z_1(\lambda_*)/Z_1(0) = 2^{L\Lambda/(2\pi)}$ with $\Lambda = \pi/a$ being the UV momentum cutoff (c.f. Sec. I C), we can write $\langle N[X_n]\rangle_{\lambda=\lambda_*} = \lambda\partial_\lambda \log(Z_1(\lambda)/Z_1(0))|_{\lambda=\lambda_*}$, which results in $\langle N[X_n]\rangle_{\lambda=\lambda_*} \propto L$. Therefore, $\langle N[X_n]\rangle_{\lambda=\lambda_*}$ does not furnish any part of the sub-leading $\mathcal{O}(1)$ contribution. This leaves us with $\langle \log \text{Det}G[X]\rangle_{\lambda=\lambda_*}$ as the only source of the sub-leading $\mathcal{O}(1)$ constant. We now perform an order by order calculation of this constant.

### A. $\mathcal{O}(\lambda^2)$

At $\mathcal{O}(\lambda^2)$, we have

$$[\langle \log \text{Det}G[X]\rangle_\lambda]_{\mathcal{O}(\lambda^2)} \equiv \mathcal{I}^{(2)} = \lambda^2 \int_{x_1<x_2} dx_1\, dx_2\, \text{Det}G[x_1, x_2] \log \text{Det}G[x_1, x_2], \tag{41}$$

where we must use the finite size propagator $G(x_1, x_2) = ((L/\pi)\sin(\pi(x_2 - x_1)/L))^{-1}$. Furthermore, at this order the normalization $(Z_1(\lambda)/Z_1(0))^{-1} = 1 - \mathcal{O}(\lambda^2)$ and so the above is the full contribution at this order. Using $\text{Det}G[x_1, x_2] = 1/d_c^2(x_1 - x_2)$ and performing a change of variables as $t = (x_1 - x_2)\pi/L$, we can simplify the above to

$$[\langle \log \text{Det}G[X]\rangle_\lambda]_{\mathcal{O}(\lambda^2)} = \lambda^2\pi \int_{\pi a/L}^{\pi[1-\frac{a}{L}]} dt\ \csc^2(t) \log\left[(\pi/L)\csc t\right] \tag{42}$$

$$= \lambda^2\pi\left(t + \cot(t) - \cot(t)\log\left((\pi/L)\csc t\right)\right)\Big|_{\pi a/L}^{\pi(1-a/L)} \tag{43}$$

$$= \pi\lambda^2\left[\pi - 2\frac{\pi a}{L}\cot\frac{\pi a}{L}\left[2\log((\pi a/L)/\sin(\pi a/L)) - 2\right]\right] \tag{44}$$

$$= -2\lambda^2\frac{L}{a} + \lambda^2\pi^2 + o(1), \tag{45}$$

where $a$ is the UV cutoff. The above contains the sub-leading constant at $\mathcal{O}(\lambda^2)$ to be $\pi^2\lambda^2$. Using $\lambda^2 = \theta^2/\pi^2 + \mathcal{O}(\theta^4)$, the $\mathcal{O}(1)$ term at this order is

$$\mathcal{I}^{(2)}|_{\mathcal{O}(1)} = \theta^2 \tag{46}$$

### B. $\mathcal{O}(\lambda^4)$

The calculation at this order is best demonstrated by regulating the theory on a periodic lattice of $L$ sites, with the lattice spacing set to unity. All positions below are integers in $\{0, \dots, L-1\}$, and we use the finite-size Majorana propagator $G_{xy} = (\pi/L)(\sin[\pi(x-y)/L])^{-1}$ for $x \neq y$ and $G_{xx} = 0$. We then define

$$W_Q(\lambda) = \log \frac{Z_Q(\lambda)}{Z_Q(0)} = \log \left[ 1 + \sum_{n=1}^{\infty} \frac{\lambda^{2n}}{(2n)!} \sum_{x_1,\dots,x_{2n}} \langle \Psi^{(Q)}_{\uparrow\downarrow}(x_1) \dots \Psi^{(Q)}_{\uparrow\downarrow}(x_{2n}) \rangle_0 \right], \tag{47}$$

where we keep the sums over $X_n = \{x_1, \dots, x_{2n}\}$ at order $n$ unordered. The above is analogous to the familiar the generating function of the *connected* contribution of the integrand in quantum field theory [18]. Therefore, we can write

$$W_Q[\lambda] = \sum_{n=1}^{\infty} \frac{\lambda^{2n}}{(2n)!} \int \sum_{x_1,\dots,x_{2n}} \langle \Psi^{(Q)}_{\uparrow\downarrow}(x_1) \dots \Psi^{(Q)}_{\uparrow\downarrow}(x_{2n}) \rangle_{0,\text{connected}} \tag{48}$$

$$= \sum_{n=1}^{\infty} \frac{\lambda^{2n}}{(2n)!} \sum_{x_1,\dots,x_{2n}} dx_i \sum_{\pi \in P_{2n}} (|\pi|-1)!(-1)^{|\pi|-1} \prod_{B\in\pi} \text{Det} G[X_B]^Q, \tag{49}$$

where $\langle \prod_{i\in B} \Psi^{(Q)}_{\uparrow\downarrow}(x_i)\rangle_0 = \langle \prod_{i\in B} \psi_{\uparrow\downarrow}(x_i)\rangle_0^{2Q} = \text{Det}G[X_B]^Q$, and where we have introduced $P_{2n}$ as non-empty partitions of $\{1, \dots, 2n\}$ with $|\pi|$ being the number of blocks in the partition $\pi$. The second equation is therfore just the expansion of cumulants in terms of products of moments as enumerated via the partitions in $P_{2n}$. From Eq. (39), we are then interested in $\lim_{Q\to1} \partial_Q W_Q[\lambda]$ which can be expressed as

$$\lim_{Q\to1} \partial_Q W_Q[\lambda] = \sum_{n=1}^{\infty} \frac{\lambda^{2n}}{(2n)!} \sum_{x_1,\dots,x_{2n}} dx_i \sum_{\pi\in P_{2n}} (|\pi|-1)!(-1)^{|\pi|-1} \left( \prod_{B\in\pi} \text{Det}G[X_B] \right) \log \left( \prod_{B\in\pi} \text{Det}G[X_B] \right), \tag{50}$$

whose $\mathcal{O}(1)$ part we will now extract at $\mathcal{O}(\lambda^4)$. At this order, we can explicitly expand the full contribution to be

$$\lim_{Q\to1} \partial_Q W_Q[\lambda] \Big|_{\mathcal{O}(\lambda^4)} \equiv \mathcal{I}^{(4)} = \frac{\lambda^4}{4!} \sum_{x_1,x_2,x_3,x_4} [\text{Det}G_{1234} \log \text{Det}G_{1234}$$
$$- \text{Det}G_{12}\text{Det}G_{34} \log \text{Det}G_{12}\text{Det}G_{34} - (13)(24) - (14)(23)], \tag{51}$$

where we use a shorthand $G_{1234}$ for the $4\times4$ covariance matrix with the BCCs placed at points $x_1, x_2, x_3, x_4$ (and similarly for the two $2\times2$ covariance matrices involving $x_i, x_j$ with $i \neq j$), and where the last two terms are the same as the second one with a different partition. The above sum can then be split into two pieces, namely, contributions where the points $x_1, x_2, x_3$, and $x_4$ are distinct and contributions when some of the lattice sites coincide.

Such coincident terms come in two types. First, when one pair of points coincide, say $x_1 = x_2 = u$, while $x_3 = v, x_4 = w$, with $u \neq v \neq w$. When this happens, the $G_{1234}$ term vanishes because the covariance matrix has repeated rows and column. Furthermore, the term involving $\text{Det}G_{12}$ is zero because $G_{uu} = 0$. The two surviving terms in the summand are equal and lead to a contribution $-2\text{Det}G_{uv}\text{Det}G_{uw} \log \text{Det}G_{uv}\text{Det}G_{uw}$. Furthermore, there is a combinatorial factor of $\binom{4}{2}2!$ coming from different ways of assigning the labels $u, v, w$ to the set of points $\{x_1, x_2, x_3, x_4\}$.

The second type of coincident contribution arises when two of the four points coincide, namely, $x_1 = x_2 = u$ and $x_3 = x_4 = v$, with $u \neq v$. Two of the four terms survive in the above summand and give $-2(\text{Det}G_{uv})^2 \log(\text{Det}G_{uv})^2$. The combinatorial factor from assigning four labels to two positions is $\binom{4}{2}$.

Separating these coincident terms out, we find

$$\mathcal{I}^{(4)} = \lambda^4(\mathcal{I}_{\text{split}} + \mathcal{I}_{\text{coincident}}), \tag{52}$$

where

$$\mathcal{I}_{\text{split}} = \sum_{x_1<\ldots x_4} [\text{Det}G_{1234}\log\text{Det}G_{1234} - \text{Det}G_{12}\text{Det}G_{34}\log\text{Det}G_{12}\text{Det}G_{34} - (13)(24) - (14)(23)] \tag{53}$$

and

$$\mathcal{I}_{\text{coincident}} = -\sum_{x_1<x_2<x_3} [\text{Det}G_{12}\text{Det}G_{13}\log\text{Det}G_{12}\text{Det}G_{13} + (12)(23) + (13)(23)] - \sum_{x_1<x_2} (\text{Det}G_{12})^2\log\text{Det}G_{12}, \tag{54}$$

and where the overall 1/4! has been canceled by the combinatorial factors obtained from restricting to ordered configurations. We can further simplify $I_{\text{coincident}}$ by expressing it in terms of the second-order contribution in Eq. (41) and the contribution with two pairs of coincident points. One can do this by first expanding the logarithms $\text{Det}G_{12}\text{Det}G_{13}\log\text{Det}G_{12}\text{Det}G_{13} = \text{Det}G_{12}\text{Det}G_{13}(\log\text{Det}G_{12} + \log\text{Det}G_{13})$. Then, separating out the translationally invariant quantity $s = \sum_{u\neq v}\text{Det}G_{uv}$ in each of those summands by accounting for further coincident terms, we can simplify to

$$\begin{aligned}\mathcal{I}_{\text{coincident}} &= -2s\sum_{x_1<x_2}\text{Det}G_{12}\log\text{Det}G_{12} + \sum_{x_1<x_2}(\text{Det}G_{12})^2\log\text{Det}G_{12} && (55)\\ &= -2s\mathcal{I}^{(2)} + R, && (56)\end{aligned}$$

where $\mathcal{I}^{(2)} = \sum_{x_1<x_2}\text{Det}G_{12}\log\text{Det}G_{12}$ denotes the coefficient of $\lambda^2$ (c.f. (41)), and $R = \sum_{x_1<x_2} dx_1dx_2(\text{Det}G_{12})^2\log\text{Det}G_{12}$. For one's own good, it is prudent to re-arrange the series at this stage into the scattering angle $\theta = \arctan(\pi\lambda)$ by using $\lambda^2 = \pi^{-2}\left(\theta^2 + \frac{2}{3}\theta^4 + \mathcal{O}(\theta^6)\right)$ and $\lambda^4 = \theta^4/\pi^4 + \mathcal{O}(\theta^6)$ to the required order. The fourth order series in the $\theta$ variable can be then expressed as

$$\mathcal{I}^{(4)} = \theta^4\left[\frac{\mathcal{I}_{\text{split}} + R}{\pi^4} + \mathcal{I}^{(2)}\left(\frac{2}{3\pi^2} - \frac{2s}{\pi^4}\right)\right]. \tag{57}$$

Several simplifications occur at this stage. First,

$$s = \frac{\pi^2}{L^2}\sum_{r=1}^{L-1}\csc^2\left(\frac{\pi r}{L}\right) = \frac{\pi^2}{L^2}\left(\frac{L^2-1}{3}\right) = \frac{\pi^2}{3}\left(1 - \frac{1}{L^2}\right), \tag{58}$$

where we have re-written the summand $s$ as a series which can be found in Ref. [19]. Substituting this into Eq. (57), we get

$$\mathcal{I}^{(4)} = \theta^4\left[\frac{\mathcal{I}_{\text{split}} + R}{\pi^4} + \frac{2\mathcal{I}^{(2)}}{3\pi^2L^2}\right], \tag{59}$$

where we see that the $2/(3\pi^2)$ in Eq. (57) cancels exactly against the constant contribution from $s$. Furthermore, since $\mathcal{I}^{(2)} = \mathcal{O}(L) + \mathcal{O}(1)$ (from Eq. (45)), the term with $\mathcal{I}^{(2)}/L^2$ also does not contribute any $\mathcal{O}(1)$ constant. Finally, we can also argue that there is no $\mathcal{O}(1)$ part in $R$ as well. Using $\text{Det}G_{12} = (\pi/L)^2\csc^2(\pi(x_1-x_2)/L)$, we write

$$\begin{aligned}R = \sum_{x_1<x_2}(\text{Det}\text{G}_{12})^2\log\text{Det}G_{12} &= \frac{L}{2}\sum_{r=1}^{L-1}(\pi/L)^4\csc^4(\pi r/L)\log(\pi/L)^2\csc(\pi r/L)^2 && (60)\\ &= -2L\sum_{r=1}^{\infty}\frac{\log r}{r^4} + \mathcal{O}(L^{-1}), && (61)\end{aligned}$$

where we have used translational invariance of the propagator. Therefore, $R$ does not contribute towards the required $\mathcal{O}(1)$ constant, leaving us with

$$\mathcal{I}^{(4)}|_{\mathcal{O}(1)} = \frac{\theta^4}{\pi^4}\mathcal{I}_{\text{split}}. \tag{62}$$

The rest of this section will be on simplifying $\mathcal{I}_{\text{split}}$ (see Eq. (53)) analytically. To begin, the determinant $\text{Det}G_{1234} = \text{Pf}(G_{1234})^2 = (G(x_{12})G(x_{34}) - G(x_{13})G(x_{24}) + G(x_{14})G(x_{23}))^2$, where $G(x_{ij}) = (\pi/L)/\sin(\pi(x_i - x_j)/L)$ is the

finite-size Majorana propagator. Using Ptolemy's trigonometric identity which states that for the chord length $s_{ij} = \sin(\phi_j - \phi_i)$, where $\phi_j$ are points $z_j = e^{2i\phi_j}$ on a circle, $s_{13}s_{24} = s_{12}s_{34} + s_{14}s_{23}$ for four ordered points defined by $0 < \phi_1 < \phi_2 < \phi_3 < \phi_4$. In our problem, $\phi_i = \pi x_i/L$ and so we can use this to simplify

$$(G(x_{12})G(x_{34}) - G(x_{13})G(x_{24}) + G(x_{14})G(x_{23}))^2 \underset{\text{Ptolemy's identity}}{=} G(x_{12})^2G(x_{34})^2 + G(x_{13})^2G(x_{24})^2 + G(x_{14})^2G(x_{23})^2 \tag{63}$$

$$= \text{Det}G_{12}\text{Det}G_{34} + \text{Det}G_{13}\text{Det}G_{24} + \text{Det}G_{14}\text{Det}G_{23}, \tag{64}$$

where we have used $G(x_{12})^2 = \text{Det}G_{12}$. Defining $P_1 = \text{Det}G_{12}\text{Det}G_{34}$, $P_2 = \text{Det}G_{13}\text{Det}G_{24}$, and $P_3 = \text{Det}G_{14}\text{Det}G_{23}$, we have

$$\mathcal{I}_{\text{split}} = \sum_{x_1<\cdots<x_4} \left[\sum_{i=1}^{3} P_i \log\left(\sum_{i=1}^{3} P_i\right) - \sum_{i=1}^{3} P_i \log P_i\right]. \tag{65}$$

Next, we introduce the conformal cross-ratio $\eta = s_{12}s_{34}/(s_{13}s_{24})$, where $s_{ij} = \sin((\pi/L)(x_i - x_j)/L)$ is the chord length. This allows us to re-write $P_{1,3}$ in terms of $P_2 = (\pi/L)^4/(s_{13}^2 s_{24}^2)$ as $P_1 = P_2/\eta^2$, and $P_3 = P_2/(1-\eta)^2$. Substituting these in the above summand, we find that a major simplification occurs and we are left with

$$\mathcal{I}_{\text{split}} = \sum_{x_1<\cdots<x_4} P_2 F(\eta), \tag{66}$$

where

$$F(\eta) = d(\eta)\log d(\eta) + 2\frac{\log\eta}{\eta^2} + 2\frac{\log(1-\eta)}{(1-\eta)^2}, \tag{67}$$

with $d(\eta) = (1 - \eta + \eta^2)^2/(\eta^2(1-\eta)^2)$. Here it is useful to use angle variables $\phi_i = \pi x_i/L$ and write the sum as

$$\mathcal{I}_{\text{split}} = \left(\frac{\pi}{L}\right)^4 \sum_{0<\phi_1<\phi_2<\phi_3<\phi_4<\pi} \frac{F(\eta)}{s_{13}^2 s_{24}^2}. \tag{68}$$

Here we introduce $\rho = \phi_1$, $a = \phi_2 - \phi_1$, $b = \phi_3 - \phi_1$, and $c = \phi_4 - \phi_1$. At this point, $0 < \rho < \pi - c$ but we can extend $\rho \in (0, \pi)$ and introducing a factor of 1/4 to account for the over-counting. This gives

$$\mathcal{I}_{\text{split}} = \frac{L}{4}\left(\frac{\pi}{L}\right)^4 \sum_{\substack{1<a<b<c<\pi \\ a,b,c\in\frac{\pi}{L}\mathbb{Z}}} \frac{F(\eta)}{\sin^2 b \sin^2(c-a)} \tag{69}$$

At this stage since all the UV collisions have been taken care of and we need to extract only the universal $\mathcal{O}(1)$ part, we can re-write the above summation as integrals in the large $L$ limit, giving

$$\mathcal{I}_{\text{split}} = \frac{\pi}{4}\int_{0<a<b<c<\pi} da\,db\,dc \frac{F(\eta)}{\sin^2 b \sin^2(c-a)}. \tag{70}$$

The next step is to use $\eta = \sin a \sin(c-b)/(\sin b \sin(c-a))$, and eliminate $a$ through a change of variables to $\eta$. After this procedure, we obtain

$$\mathcal{I}_{\text{split}} = \frac{\pi}{4}\int_{0<b<c<\pi}\int d\eta \frac{db\,dc}{\sin b \sin(c-b)\sin c} F(\eta). \tag{71}$$

Re-introducing the UV cutoff $\epsilon = \pi/L$, we require these angles to satisfy $b > \epsilon, c - b > \epsilon$, and $\pi - c > \epsilon$, leading to the ranges $2\epsilon < c < \pi - \epsilon$ and $\epsilon < b < c - \epsilon$ for $c$ and $b$ respectively. Therefore,

$$\mathcal{I}_{\text{split}} = \frac{\pi}{4}\int_0^1 F(\eta)d\eta \int_{2\epsilon}^{\pi-\epsilon} \frac{dc}{\sin c}\int_{\epsilon}^{c-\epsilon} \frac{db}{\sin b \sin(c-b)}, \tag{72}$$

where the integrals have been factorized since any overlapping pieces do not give an $\mathcal{O}(1)$ contribution. Performing these integrals results in

$$\mathcal{I}_{\text{split}} = \left(\int_0^1 F(\eta)d\eta\right)\left[\frac{3\pi\log 2}{2\epsilon} - \frac{\pi^2}{2} + \frac{\pi}{2}\left(\frac{3}{2} - \log 2\right)\epsilon + \mathcal{O}(\epsilon^3)\right]. \tag{73}$$

Using $\int_0^1 F(\eta)d\eta = 2\sqrt{3}\pi$, we have

$$\mathcal{I}^4|_{\mathcal{O}(1)} = \frac{\theta^4}{\pi^4}\left(-\sqrt{3}\pi^3\right) = -\frac{\sqrt{3}}{\pi}\theta^4. \tag{74}$$

Using Eq. (40), (62), (74), we have

$$s_{\text{eff}} = \log g_{\text{eff}}(\theta_*) = \left[\theta_*^2 - \frac{\sqrt{3}}{\pi}\theta^4 + \mathcal{O}(\theta_*^6)\right] - \log 2, \tag{75}$$

where the $\log 2$ comes from the universal sub-leading constant in $\log p_\uparrow$, which is simply $2\log g_\uparrow$. Up to $\mathcal{O}(\theta^4)$, we therefore have $s_{\text{eff}} = (\pi/4)^2 - \sqrt{3}\pi^3/(4^4) - \log 2 = -0.2860798999$.

## III. MEASUREMENT-ALTERED CORRELATORS

### A. Setup

In this section we will present a detailed calculation of the measurement-averaged spin-spin correlators upon weakly measuring the $\sigma^z$ operator (or in the field theory, the $\sigma$ field) with measurement strength $\gamma$ (see Sec. I). In particular, we will be interested in the moments of connected correlators

$$\mathcal{C}_k(x) = \overline{[\delta\sigma_0^z\delta\sigma_x^z]^k} = \sum_m p_m \langle\delta\sigma_0^z\delta\sigma_x^z\rangle_m^k, \tag{76}$$

where $\delta\sigma^z = \sigma^z - \langle\sigma^z\rangle$. In the continuum, the lattice operator can be expanded in terms of the scaling fields of the critical theory, that is, $\sigma_x^z \simeq a^{1/8}\sigma(x)$ , where $a$ is the UV cutoff and $x = ja$ for some $j \in \mathbb{Z}^{>0}$. As discussed in Sec. I C, the Born-weight of the coarse grained measurement outcomes enumerated by the set of BCC insertions $X_n$ is $\langle\Psi[X_n]\rangle_0 = \langle\prod_{i=1}^{2n}\psi^{(1)}_{\uparrow\downarrow}(x_i)\psi^{(2)}_{\uparrow\downarrow}(x_i)\rangle_0 = \text{Det}\, G_n$, with $n \in \mathbb{Z}$. Accordingly, we can directly write the coarse-grained form of Eq. (76) as

$$\mathcal{C}_k(x) = \frac{Z_1(0)}{Z_1(\lambda)}\sum_{n=0}^{\infty}\lambda^{2n}\int_{x_1<\cdots<x_{2n}} dX_n\ \text{Det}G[X_n]\left[\frac{\langle\Psi[X_n]\sigma_0^z\sigma_x^z\rangle_0}{\text{Det}G[X_n]} - \frac{\langle\Psi[X_n]\sigma_0^z\rangle_0}{\text{Det}G[X_n]}\frac{\langle\Psi[X_n]\sigma_x^z\rangle_0}{\text{Det}G[X_n]}\right]^k, \tag{77}$$

where $dX_n = \prod_{i=1}^{2n} dx_i$ and where we have replaced $m \to \mathcal{X} = \{X_n|\ n \in \mathbb{Z}\}$ and $p_m \to \text{Det}G[X_n]$ (up to normalization $Z_1(0)/Z_1(\lambda)$). The above expression also follows from the general BCFT recipe to evaluate such averaged correlators proposed in Ref. [11]. *At* the $\gamma \to \infty$ fixed point, the above correlators vanish identically since we are effectively projecting out the lattice $\sigma^z$ degrees of freedom. Therefore, any non-trivial decay of the above correlator must arise from leading corrections when one shifts away from this fixed point and allows the spins to fluctuate. Such contributions can be included by introducing irrelevant perturbations to the $\gamma \to \infty$ fixed points, as per the allowed operator spectrum in the fixed boundary sector encoded in the partition function $Z_{\uparrow\uparrow} = \chi_{\mathbb{I}}(q) \underset{q\to 0}{=} q^{-1/48}(1+q^2+q^3+2q^4+\dots)$. These irrelevant corrections can be systematically restored in the spin field $\sigma(x)$ through its bulk-boundary OPE with the fixed boundary condition [20]:

$$\sigma(x,\epsilon)|_{s=+1,-1} \underset{\epsilon(\gamma)\to 0}{\sim} \frac{1}{\epsilon(\gamma)^{1/8}} s\left[C_{\mathbb{I}}\mathbb{I} + C_T\epsilon(\gamma)^2 T(x) + \dots\right], \tag{78}$$

where $\epsilon(\gamma)$ is the distance from the boundary and is parametrized by the parameter $\gamma$ with $\epsilon(\gamma) \to 0$ as $\gamma \to \infty$, $T(x)$ is the boundary stress-energy tensor and the ellipses represent further (boundary) descendants of the identity Virasoro spectrum which appears at a fixed boundary. $C_{\mathbb{I},T}$ are constants which are universal up to normalization conventions [20] but will be unimportant for subsequent analysis. The sign $s = +1, -1$ is set by the direction of the fixed boundary with $s = +1(-1)$ for the $|\uparrow\rangle$ ($|\downarrow\rangle$) boundary condition. Using the above OPE in Eq. (77), it is clear that the contribution coming from $\mathbb{I}$ in Eq. (78) is zero since the connected and disconnected parts are identical. This is equivalent to the fact that at the $\gamma \to \infty$ fixed point, there are no non-trivial (connected) correlations. The first non-trivial contribution arises from fluctuations away from this fixed point in the form of the stress energy tensor in (78) and is given as

$$\mathcal{C}_k(x-y) = \sum_{n=0}^{\infty}\lambda^{2n}\int dX_n\ \text{Det}G[X_n](s(x)s(y))^k\left[\frac{\langle\Psi[X_n]T(y)T(x)\rangle_0}{\text{Det}G[X_n]} - \frac{\langle\Psi[X_n]T(y)\rangle_0}{\text{Det}G[X_n]}\frac{\langle\Psi[X_n]T(x)\rangle_0}{\text{Det}G[X_n]}\right]^k, \tag{79}$$

where we omit overall constants and replace $0 \to y$ in order to help clarity of presentation in the coming sections. $s(x)$ is the sign of the local boundary condition at $x$ and depends on the overall configuration $X_n$.

### B. Evaluation of Moments

We now determine the long-distance scaling of $\mathcal{C}_k(x-y)$ for arbitrary positive integer $k$. While the integral in Eq. (79) is hard to evaluate in general, it will be sufficient to isolate the most slowly decaying contribution (in $|x-y|$) to the connected stress-energy correlator for a given configuration $X_n$.

We begin by stating two boundary OPE relations that we will use frequently, namely,

$$T(x)T(y) \sim \frac{c/2}{(x-y)^4} + \frac{2T(y)}{(x-y)^2} + \frac{\partial_y T(y)}{x-y} + \cdots, \tag{80}$$

and

$$\langle T(x)\Psi[X_n]\rangle_0 = \sum_{i=1}^{2n}\left[\frac{1}{(x-x_i)^2} + \frac{1}{x-x_i}\partial_{x_i}\right]\langle\Psi[X_n]\rangle_0 . \tag{81}$$

Applying the above OPE relations recursively, we can simplify the connected part of Eq. (79) as

$$\langle\Psi[X_n]T(x)T(y)\rangle_0 = \underbrace{\frac{c/2}{(x-y)^4}\langle\Psi[X_n]\rangle_0}_{A} + \underbrace{\left[\frac{2}{(x-y)^2} + \frac{1}{x-y}\partial_y\right]\langle\Psi[X_n]T(y)\rangle_0}_{B} + \underbrace{\sum_{i,k=1}^{2n}\partial_{x_i}\left[\frac{1}{(x-x_i)}\partial_{x_k}\left[\frac{1}{(y-x_k)}\langle\Psi[X_n]\rangle_0\right]\right]}_{C}. \tag{82}$$

$A$ and $B$ include the singular contributions when $T(x)$ fuses with $T(y)$ while $C$ is the singular contribution when $x, y$ approach two points in $x_i, x_k \in X_n$. $A$ contributes at least $\sim (x-y)^{-4}$ and will therefore be an overall sub-leading contribution to $\mathcal{C}_k(x-y)$. Next, we consider $C$, which can be expanded as

$$C = \sum_{i,k}\frac{\langle\Psi[X_n]\rangle_0}{(x-x_i)^2(y-x_k)^2} + \sum_{i,k}\frac{1}{(x-x_i)^2}\frac{\partial_{x_k}\langle\Psi[X_n]\rangle_0}{y-x_k} + \sum_{i,k}\frac{\partial_{x_i}}{x-x_i}\left[\frac{\langle\Psi[X_n]\rangle_0}{(y-x_k)^2}\right] + \sum_{i,k}\frac{\partial_{x_i}}{x-x_i}\left[\frac{\partial_{x_k}\langle\Psi[X_n]\rangle_0}{y-x_k}\right]. \tag{83}$$

While the above looks complicated, one can use OPE relation Eq. (81) to simplify the disconnected combination $\langle\Psi[X_n]T(y)\rangle_0\langle\Psi[X_n]T(x)\rangle_0$. Upon doing so, one finds that the first two terms in $C$ cancel exactly with parts of the disconnected correlator and hence do not contribute. Labeling these terms $C_{\text{disc},1}$, we have

$$C - C_{\text{disc},1} = \sum_{i,k}\frac{\partial_{x_i}}{x-x_i}\left[\frac{\langle\Psi[X_n]\rangle_0}{(y-x_k)^2}\right] + \sum_{i,k}\frac{\partial_{x_i}}{x-x_i}\left[\frac{\partial_{x_k}\langle\Psi[X_n]\rangle_0}{y-x_k}\right]. \tag{84}$$

Further expanding the derivatives in $C - C_{\text{disc},1}$, we get

$$C - C_{\text{disc},1} = \sum_{i,k}\frac{1}{x-x_i}\left[\frac{\partial_{x_i}\langle\Psi[X_n]\rangle_0}{(y-x_k)^2} + \frac{2\delta_{ik}\langle\Psi[X_n]\rangle_0}{(y-x_k)^3}\right] + \sum_{i,k}\frac{1}{x-x_i}\left[\frac{\partial_{x_i}\partial_{x_k}\langle\Psi[X_n]\rangle_0}{y-x_k} + \frac{\delta_{ik}\partial_{x_k}\langle\Psi[X_n]\rangle_0}{(y-x_k)^2}\right]. \tag{85}$$

Here too, the first term cancels against a term in the disconnected piece $\langle\Psi[X_n]T(y)\rangle_0\langle\Psi[X_n]T(x)\rangle_0$, leaving behind

$$C_{\text{connected}} = C - C_{\text{disc},1} - C_{\text{disc},2} = \sum_{i}\frac{1}{x-x_i}\frac{2\langle\Psi[X_n]\rangle_0}{(y-x_i)^3} + \sum_{i,k}\frac{1}{x-x_i}\left[\frac{\partial_{x_i}\partial_{x_k}\langle\Psi[X_n]\rangle_0}{y-x_k} + \frac{\delta_{ik}\partial_{x_k}\langle\Psi[X_n]\rangle_0}{(y-x_k)^2}\right], \tag{86}$$

where we have labeled the disconnected piece in Eq. (85) as $C_{\text{disc},2}$. The leftover pieces are now genuine connected contributions from $C$ and are therefore labeled as $C_{\text{connected}}$. Next, we apply the OPE (81) to the piece $B$ in Eq. (82), giving us

$$B = \frac{2}{(x-y)^2}\sum_i\left[\frac{1}{(y-x_i)^2} + \frac{\partial_{x_i}}{(y-x_i)}\right]\langle\Psi[X_n]\rangle_0 + \frac{1}{(x-y)}\sum_k\left[\frac{-2}{(y-x_k)^3} + \frac{-\partial_{x_k}}{(y-x_k)^2}\right]\langle\Psi[X_n]\rangle_0. \tag{87}$$

The above are all genuine connected pieces since they stem from the $x \to y$ singularity of stress energy tensors $T(x)$ and $T(y)$. At the outset, it seems as though the slowest connected contribution stem from the terms in $B$ that scale as $(x-y)^{-1}$ (from the latter terms above). Indeed, if we consider $x_k = y + a$ in these terms, where $a$ is the UV cutoff, we obtain $(x-y)^{-1}\sum_k(2a^{-3} - a^{-2}\partial_k)\langle\Psi[X_n]\rangle_0$, which scales as $(x-y)^{-1}$. However, for every

such singular piece in $B$, there's a corresponding piece in $C_{\text{connected}}$ that cancels it out. For instance, if we similarly consider taking a BCC at $x_i = y + a$ in the first and third terms in $C_{\text{connected}}$, we find the singular contribution $(x-y)^{-1}\sum_i\left(-2a^{-3}+a^{-2}\partial_i\right)\langle\Psi[X_n]\rangle_0$ which indeed is the same singular contribution as $B$ but with an opposite sign. Therefore, such singularities when BCCs approach either $x$ or $y$, which contribute towards a $(x-y)^{-1}$ decay always cancel between $B$ and $C$. Taking these BCCs to then lie at a distance that scales with $|x-y|$, and setting for example $|y-x_i| = u|x-y|$, with $u \sim \mathcal{O}(1)$, will lead to higher powers in $(x-y)$. Accordingly, the next leading contribution is $\sim (x-y)^{-2}$ stemming from several terms in both $C_{\text{connected}}$ and $B$. Here once again one may analyze whether such contributions cancel between $B$ and $C_{\text{connected}}$ or not, which is a rather tedious task. We instead consider isolate the contribution from the double derivative term in $C_{\text{connected}}$ whose $(x-y)^{-2}$ scaling stems from a fundamentally different mechanism from other terms in $C_{\text{connected}}$ and $B$. It occurs when *both* $x$ and $y$ approach distinct BCC positions $x_i$ and $x_k$ with $i \neq k$ [the $i = k$ piece exists in principle but does not contribute to this mechanism since we keep $|x-y| \gg a$]. The leading $(x-y)^{-2}$ from this term does not then arise in any other terms and therefore this always survives. Combining this double derivative piece from $C_{\text{connected}}$ and a corresponding piece in the disconnected part $\langle\Psi[X_n]T(y)\rangle_0\langle\Psi[X_n]T(x)\rangle_0$, we write

$$\mathcal{C}_k(x-y) \simeq \sum_{n=0}^{\infty}\lambda^{2n}\int dX_n\ \mathrm{Det}G[X_n](s(x)s(y))^k\Gamma(x,y)^k, \tag{88}$$

where

$$\Gamma(x-y) = \sum_{i\neq k}\frac{1}{(x-x_i)(y-x_k)}\left[\frac{\partial_i\partial_k\langle\Psi[X_n]\rangle_0}{\langle\Psi[X_n]\rangle_0} - \frac{\partial_i\langle\Psi[X_n]\rangle_0}{\langle\Psi[X_n]\rangle_0}\frac{\partial_k\langle\Psi[X_n]\rangle_0}{\langle\Psi[X_n]\rangle_0}\right] \tag{89}$$

$$\underset{\substack{y\to x_k\\ x\to x_i}}{\simeq} a^{-2}\sum_{i\neq k}\left[\frac{\partial_x\partial_y\langle\Psi[X_n]\rangle_0}{\langle\Psi[X_n]\rangle_0} - \frac{\partial_i\langle\Psi[X_n]\rangle_0}{\langle\Psi[X_n]\rangle_0}\frac{\partial_k\langle\Psi[X_n]\rangle_0}{\langle\Psi[X_n]\rangle_0}\right]\Bigg|_{x_k\simeq y, x_i\simeq x} \tag{90}$$

$$= a^{-2}\sum_{i\neq k}\partial_x\partial_y\log\left(\langle\Psi[X_n|x_k=y, x_i=x]\rangle_0\right). \tag{91}$$

Since the scaling dimension of $\partial_x\partial_y\log\left(\langle\Psi[X_n|x_k=y,x_i=x]\rangle_0\right)$ is 2, we can write in the scaling limit $a/|x-y|\to 0$,

$$\Gamma(x,y) = a^{-2}\sum_{i,k}(x-y)^{-2}\mathcal{F}_{i,k}\left(\left\{\frac{|x_j-y|}{|x-y|}\right\}_{j\neq i,k}\right), \tag{92}$$

where $\mathcal{F}$ consists of dimensionless ratios of the length-scales in the problem. Substituting the above scaling in Eq. (88), we see that for even $k = 2m$, $m \in \mathbb{Z}$, the sign terms $(s(x)s(y))^{2m} = 1$ and the correlator scales as $\mathcal{C}_k(x-y) \sim (x-y)^{-4m}$ since the remaining combination is dimensionless. For odd $k$, the remaining integral scales like $\sim \int dX_n \mathrm{Det}G[X_n]s(x)s(y)$. This is because the function $\mathcal{F}_{i,k}$ only alters the BCC configuration weights by an $\mathcal{O}(1)$ factor and is not expected to alter the scaling in $(x-y)$. The scaling of the remaining integral is $\sim (x-y)^{-1/4}$ since it is same spin-spin correlator we studied in Sec. I C, where the spin field $\sigma(x)$ was also reduced to a sign depending on the local boundary condition. Putting everything together then gives

$$\mathcal{C}_k(x-y) \sim \begin{cases}|x-y|^{-2k} & k\in 2\mathbb{Z}\\ |x-y|^{-2k-1/4} & k \in 2\mathbb{Z}+1.\end{cases} \tag{93}$$


\end{document}